\documentclass{aa}

\usepackage{graphicx}
\usepackage{longtable}
\usepackage{soul}
\usepackage{txfonts}
\usepackage{placeins}

\usepackage[colorlinks=true, allcolors=blue]{hyperref}

\usepackage{xcolor}
\DeclareUnicodeCharacter{0301}{\'{e}}
\DeclareUnicodeCharacter{0301}{\'{i}}
\DeclareUnicodeCharacter{2212}{-}

\usepackage{colortbl}
\usepackage{multirow}
\usepackage{comment}
\usepackage{subcaption}

\begin{document}

   \title{Probing accretion and stellar properties in the Orion Nebula with VLT/X-Shooter\thanks{Based on observations collected at the European Southern Observatory under ESO programmes 0108.C-0919 and 0114.D-0441}}
   \authorrunning{Piscarreta et al.}

   \author{L. Piscarreta
          \inst{1}, G. Beccari\inst{1}, R. A. B. Claes\inst{1}, C. F. Manara\inst{1}, H. M. J. Boffin\inst{1}, T. Jerabkova\inst{2}, \\B. Ercolano\inst{3,4,5}, A. Natta\inst{6}, and S. E. van Terwisga\inst{7}
          }

   \institute{European Southern Observatory, Karl-Schwarzschild-Strasse 2, 85748 Garching bei München, Germany\\
              \email{lara.alvopiscarreta@eso.org}
         \and
         Institute of Theoretical Physics and Astrophysics, Masaryk University, Kotlářská 2, Brno, 611 37, Czech Republic
         \and
         University Observatory, Faculty of Physics, Ludwig-Maximilians-Universität München, Scheinerstr 1, D-81679 Munich, Germany
         \and
         Exzellenzcluster “Origins”, Boltzmannstr. 2, D-85748 Garching, Germany
         \and
         Max-Planck-Institut für Extraterrestrische Physik, Giessenbachstra.e 1, 85748 Garching, Germany
         \and
         School of Cosmic Physics, Dublin Institute for Advanced Studies, 31 Fitzwilliam Place, Dublin 2, Ireland
         \and
         Space Research Institute, Austrian Academy of Sciences, Schmiedlstr. 6, 8042, Graz, Austria
             }

   \date{Received 13 May 2025; Accepted 6 September 2025}

  \abstract
   {Multiple photometric studies have reported the presence of seemingly older accreting pre-main sequence stars (PMS) in optical colour-magnitude diagrams (CMDs). These sources appear bluer than the majority of cluster members, leading to older isochronal age estimates.} 
   {We investigate this phenomenon in the Orion Nebula, which harbors a subset of stars that show infrared excess detected by \textit{Spitzer} (probing the presence of protoplanetary discs) and H$\alpha$ excess emission (tracing ongoing mass accretion), yet display significantly older isochronal ages ($\gtrsim$10 Myr) compared to the bulk population ($\sim$1$-$3 Myr) in the $r, (r-i)$ CMD. This raises the question of whether these stars are truly older or whether their photometric properties are affected by observational biases or other physical processes.}
    {We perform a detailed spectroscopic analysis of 40 Orion Nebula stars using VLT/X-Shooter, covering CMD-based isochronal ages from 1 to over 30 Myr. We derive extinction values, stellar properties, and accretion parameters by modeling the ultraviolet excess emission through a multicomponent fitting procedure. The sample spans spectral types from M4.5 up to K6, and masses in the range $\sim$0.1$-$0.8 M$_{\odot}$.}
    {We demonstrate that, when extinction and, most importantly, accretion effects are accurately constrained, the stellar luminosity and effective temperature of the majority of the seemingly old stars become consistent with a younger population ($\sim$1$-$5 Myr). This is supported by strong lithium absorption (EW$_{\rm Li} \gtrsim$ 400 m$\AA$), which corroborates their youth, and by the accretion-to-stellar luminosity ratios ($L_{\rm acc}/L_{\star}$) typical for young, accreting stars. Three of these sources, however, remain old even after our analysis, despite showing signatures consistent with ongoing accretion from a protoplanetary disc. More generally, our analysis indicates that excess continuum emission from accretion shocks affects the placement of PMS stars in the CMD, displacing sources towards bluer optical colours.}
    {This study highlights the critical role of accretion in shaping the stellar properties estimates (including age) derived from optical CMDs and emphasizes the need to carefully account for accretion effects when interpreting age distributions in star-forming regions. Understanding these biases is essential for accurately constraining the early evolution of PMS stars.}


   \maketitle

\section{Introduction}

The evolution of protoplanetary discs around pre-main sequence (PMS) stars is a fundamental process in star and planet formation. These discs provide the material for planetary systems, and their dispersal timescale sets a critical time ceiling for planet formation to occur (e.g., \citealt{Alexander_2014}). During the early stages of disc evolution, accretion of material onto the central star plays a key role in PMS evolution. Understanding how accretion depends on stellar properties, such as mass and age, as well as on environmental conditions, is crucial for constraining the physical mechanisms driving early stellar evolution (e.g., \citealt{coleman_2022}).

To study disc evolution, it is first necessary to reliably identify disc-bearing stars across different stellar populations. This is commonly done by detecting an infrared (IR) excess in the spectral energy distributions, originating from warm dust in the disc \citep{Lada_1984,Lada_1987,Greene_1994}. Another key signature is ongoing accretion, which can be detected through excess emission in the ultraviolet (UV; e.g., \citealt{Calvet_Gullbring_1998,Herczeg_Hillenbrand_2008}) or strong emission lines such as H$\alpha$ (e.g., \citealt{White_2003, BarradoNavascues_2003}).
Using these observational diagnostics, studies of nearby star-forming regions (SFRs) have shown that 1) the fraction of stars with discs decreases approximately exponentially, with typical reported disc lifetimes of $\sim$2$-$3 Myr when based on near-infrared (NIR) excess \citep[e.g.,][]{Hernandez_2007,WilliamsCieza_2011,Ribas_2015}; and 2) accretion declines on a comparable or shorter timescale, with the fraction of accreting PMS stars dropping to only 2$-$3\% by 10 Myr \citep[e.g.,][]{Fedele_2010, Delfini_2025}.

However, PMS stars exhibiting significant accretion signatures at ages $\gtrsim$20 Myr have been reported in different spectroscopic and photometric studies \citep[e.g.,][]{Murphy_2018,Silveberg_2020,Biazzo_2017,DeMarchi_2024}. Various processes such as external gas infall, the existence of discs showing both debris-like and primordial disc features, and interactions between stars or between stars and discs have been suggested as possible explanations for these so-called "Peter Pan" discs (e.g., \citealt{Scicluna_2014,Silveberg_2020}), highlighting the complex interplay between internal and external processes that shape disc evolution. Interpreting these cases is further complicated by the uncertainties in age determinations since photometric ages derived from colour-magnitude diagrams (CMDs) can be affected by extinction, variability, episodic accretion, and disc inclination \citep[e.g.,][]{Baraffe_2010,Soderblom_2014,Jeffries_2017, DeMarchi_2013}. On the other hand, spectroscopic age determinations can be influenced by strong accretion, which veils photospheric features, complicating spectral classification and thus effective temperature determination \citep{Fernandez_2001,Huelamo_2010}, and edge-on disc geometries that obscure the stellar photosphere making the stars appear underluminous, hence colder and older than they truly are \citep[e.g.][]{Alcala_2014}.

The Orion Nebula, as the closest ($\sim$385$-$395 pc; \citealt{Kounkel_2022}) massive  ($\sim$1500 M$_{\odot}$, \citealt{Kroupa_2018}) SFR, provides a rich environment for studying PMS evolution in a dynamically complex setting. It hosts a large and diverse population of low-mass YSOs, while also containing massive O- and B-type stars, such as those in the Trapezium Cluster. While the bulk population has a mean age of 2.2 Myr, with a scatter of a few Myrs \citep{Reggiani_2011}, some sources appear significantly older in the CMD, despite displaying clear disc and accretion signatures. 

It is still highly debated whether the apparent age spread in the Orion Nebula is primarily attributed to observational biases, suggesting a coeval population (e.g., \citealt{Jeffries_2011}), or whether it reflects a real distribution of stellar ages (e.g., \citealt{DaRio_2010,DaRio_2016}), potentially indicative of extended or multiple episodes of star formation (e.g., \citealt{Beccari_2017,Jerabkova_2019}). Lithium depletion studies by \citet{Palla_2005,Palla_2007} revealed a small number of Orion Nebula members with significantly depleted lithium, pointing to ages older than 10 Myr, well beyond the age typically assumed for this region’s population ($\sim$1$-$3 Myr). However, it remains possible that some sources that appear older in CMDs are not intrinsically older, but rather affected by observational or physical effects that can bias age estimates.

In this paper, we perform a homogeneous spectroscopic study of 40 PMS stars hosting a disc in the Orion Nebula. The aim of this work is to study accretion properties as a function of the host star parameters, with a particular focus on a subset of disc-bearing sources with ongoing accretion and seemingly older isochronal ages than those expected for Orion Nebula sources.

\section{Observations and data reduction}
\label{sec:data}

\begin{figure}[h!]
\centering
    \includegraphics[width=0.92\columnwidth]{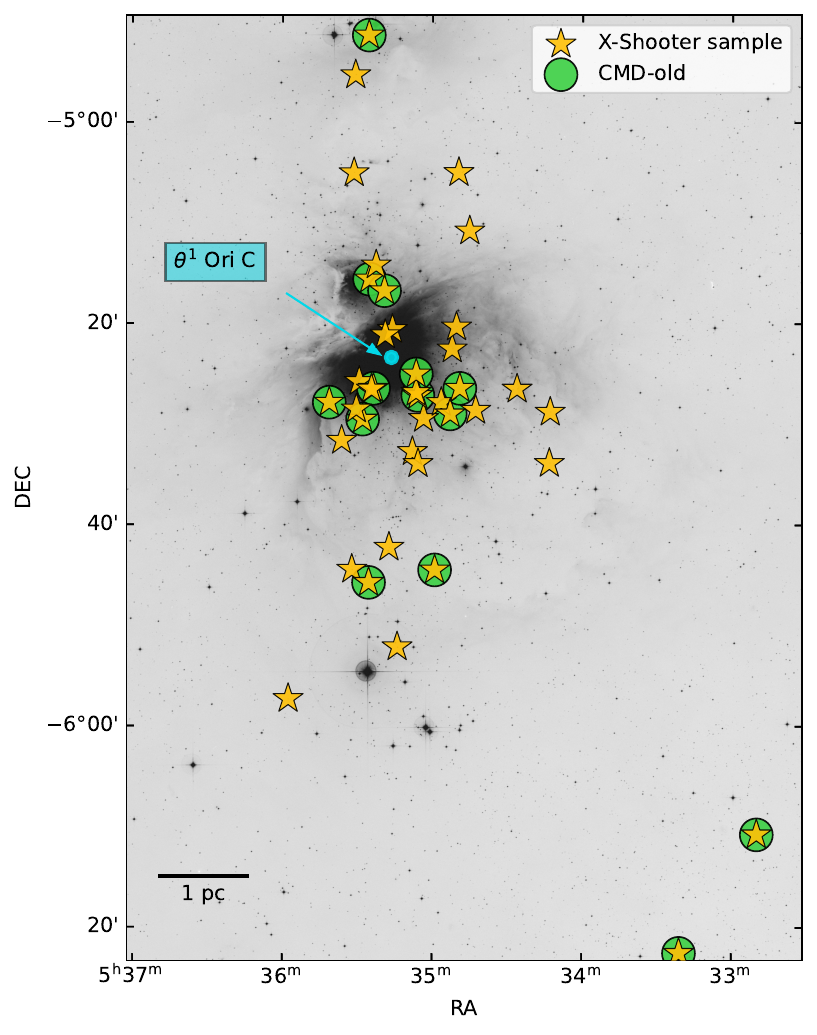}
\caption{Position of the targets in the sky. The X-Shooter sample is marked as yellow stars, and sources with isochronal ages $\gtrsim$ 10 Myr in the CMD are pinpointed with green circles. The position of $\theta^{1}$ Ori C is marked with a cyan circle. The background is a DSS2-infrared image \citep{McLean_2000}.}
\label{fig:ON_fov}
\end{figure}

We observed 40 disc-bearing PMS stars in the Orion Nebula with ESO VLT/X-Shooter at the Very Large Telescope (VLT) in Paranal, Chile. X-Shooter \citep{Vernet_2011} is an intermediate-resolution spectrograph that covers, simultaneously, a broad wavelength range, consisting of three independent arms: UVB (300$-$550 nm), VIS (550$-$1000 nm), and NIR (1000$-$2480 nm). The observations were carried out in Service Mode (SM), with the first program conducted between October 2021 and January 2022 (program ID 0108.C-0919, PI Beccari), providing 33 spectra. We include in this study seven spectra taken with X-Shooter between October 2024 and February 2025 as part of the SM program 0114.D-0441 (PI Piscarreta). These 40 spectra constitute a well-defined and self-contained sample, selected to investigate accretion properties in the Orion Nebula and, in particular, to characterise a subset of seemingly old accreting PMS stars. The targets have the following properties:
\begin{enumerate}
    \item Classified as “Disk” according to \textit{Spitzer} \citep{Megeath_2012};
    \item Located within 1 degree from the centre of the Trapezium cluster and at an angular separation greater than $\sim$2.34 arcminutes ($\sim$0.28 pc in projection) from $\theta^{1}$ Ori C — the O6V primary ionising source in the Orion Nebula Cluster (\citealt{ODell_2017}; see Figure \ref{fig:ON_fov});
    \item Have parallaxes between $\sim$2.15$-$2.92 mas (i.e., distance $\sim$340$-$465 pc) according to \textit{Gaia} DR3 \citep{Gaia_2016,Gaia_2023};
    \item Selected based on their location in the $r$–H$\alpha$ vs. $r$–$i$ colour-colour diagram using OmegaCAM photometry from the Accretion Discs in H$\alpha$ with OmegaCAM (ADHOC) survey (program IDs 096.C-0730(A) and 098.C-0850(A), PI Beccari). Most of the sample (32 sources out of 40) shows strong H$\alpha$ excess consistent with ongoing accretion (EW$_{\rm H\alpha}>$20 $\AA$), while the remaining 8 sources lie closer to the locus of non-accreting stars, with weaker or marginal H$\alpha$ excess.
\end{enumerate}

\begin{figure}[h!]
\centering
    \includegraphics[width=\columnwidth]{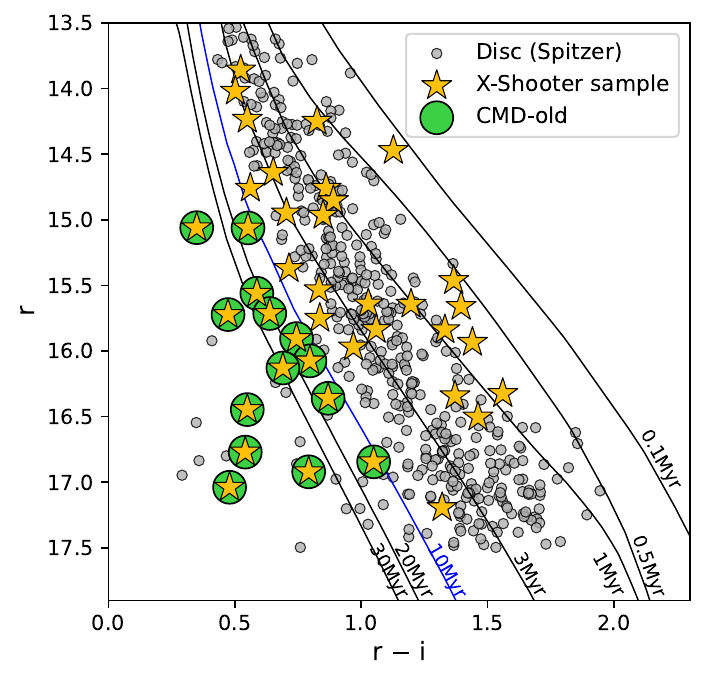}

\caption{\textit{r, (r$-$i)} OmegaCAM colour-magnitude diagram for the X-Shooter sample (yellow stars) with sources $\gtrsim$10 Myr (isochrone highlighted in blue) pinpointed with green circles (CMD$-$old). Sources with a disc within 1 degree from the center of the Orion Nebula Cluster according to \textit{Spitzer} photometry \citep{Megeath_2012} are displayed as dark gray circles. Only sources brighter than 17.5 mag are shown. The solid lines show the location of 0.1, 0.5, 1, 3, 10, 20, and 30 Myr PISA PMS isochrones \citep{PISA_isochrones}.}
\label{fig:cmd}       
\end{figure}

We show in Figure \ref{fig:cmd} the position of the objects in the $r$ vs. $r-i$ CMD. While most stars have isochronal ages between $\sim1$ and $\sim3$Myr, consistent with the Orion Nebula population, a subset appears significantly older. Specifically, 14 out of the 40 stars in our sample exhibit both isochronal ages $\gtrsim$ 10 Myr and H$\alpha$ excess emission (highlighted with green circles in Figure \ref{fig:cmd}). We refer to this group as the "CMD$-$old" population throughout this work. The remaining 26 stars, which span a similar mass range ($\sim$0.1$-$0.8 M$_{\odot}$), show younger ages and serve as a representative sample of the Orion Nebula bulk population. The OmegaCAM photometry for the full sample is provided in Table \ref{tab:omegacam_info}.

The targets were observed in slit-nodding mode adopting a ABBA cycle strategy with slit widths of 1.3'' for the UVB arm and 0.9'' for the VIS and NIR arms in both programs. This ensured spectral resolutions of R$\sim$4100, 8900, and 5600 in the UVB, VIS, and NIR arms, respectively. We adopted for each offset position exposure times equal to 254s, 160s, and 320s in the UVB, VIS, and NIR arms, respectively, for stars with magnitudes $r_{\rm AB}<15$ mag. Stars with 15 $< r_{\rm AB} <$17.5 mag were observed using 694s, 600s, and 700s exposure times for each offset position in the UVB, VIS, and NIR arms, respectively. 

The median seeing during the observations was 0.88" with a standard deviation of 0.22". The observations were carried out under relaxed weather constraints, meaning that while some exposures were obtained in photometric or clear skies, most of the spectra were acquired with thin cirrus in the sky. We obtained an additional short exposure in stare mode using a 5.0"-slit in all three arms immediately after each science exposure, in order to guarantee accurate flux calibration. The spectra obtained using a 5.0"-slit do not suffer from flux losses and thus act as a reference to properly flux-calibrate the narrow-slit observations.

The data reduction was performed using the X-Shooter pipeline \citep{xshooter_pipeline} v3.6.3 within the \textit{EsoReflex} environment \citep{esoreflex}. The pipeline applies standard reduction steps, including bias and dark subtraction, flat-field correction, order tracing and merging, wavelength calibration, sky subtraction, 1D spectrum extraction, and flux calibration using standard stars observed on the same night as the science observations. The spectra from the three X-Shooter arms were reduced independently. 

The presence of strong and variable nebular emission lines in the Orion Nebula introduces significant sky variability, making the sky subtraction performed by the X-Shooter pipeline not ideal for some sources. In some cases, this led to artifacts in the extracted one-dimensional spectra, such as artificially deep central absorptions in emission lines (e.g., Balmer lines), likely caused by over-subtraction or sky saturation at those wavelengths. While sky subtraction can be optimized by adjusting parameters in the \textit{EsoReflex} environment, we found it useful to complement this approach with \texttt{Jdaviz/Specviz2D} \footnote{\url{https://jdaviz.readthedocs.io}}. This package allowed us to interactively inspect the rectified and order-merged 2D data product obtained from the pipeline and manually assess how different background selections affected the final extracted line profiles. Subtracting the average background from a visually refined sky selection led to cleaner extracted spectra.

The telluric correction was performed with \textit{Molecfit} \citep{Smette_2015,Kausch_2015} v4.3.1 for the VIS and NIR arms. 
Finally, we rescaled the narrow-slit observations to their respective wide slit ones to obtain the final flux calibration. The procedure is the same as presented in \citet{Manara_2021}. The scaling factor is computed as the median ratio between the wide-slit and narrow-slit observations (considering spectral regions outside the telluric absorption bands), or it can vary linearly with wavelength. We used the first approach for the UVB arm and the second for the VIS and NIR arms.

\section{Analysis}\label{sec:FRAPPE}
\subsection{Multicomponent fitting procedure}
Obtaining accurate stellar and accretion properties in YSOs is challenging due to the combined effects of extinction and geometry. Circumstellar material, such as protoplanetary discs and any residual envelope material, contributes to extinction \citep{Natta_2006,Manara_2013a}, and disentangling these components is not straightforward. Moreover, the amount of extinction also depends on the system's inclination, which further complicates the determination of the star's intrinsic photospheric emission.

In addition to these geometric and environmental factors, the accretion process itself introduces further complexity. According to the magnetospheric accretion model \citep[see review by][]{Hartmann_2016}, the process of material being funneled from the inner edge of the disc onto the stellar surface along the magnetic field lines results in accretion shocks that generate excess continuum emission (mainly in the Balmer continuum, $\lambda \lesssim$ 364.6 nm, and the Paschen continuum 364.6$\lesssim \lambda \lesssim$ 820 nm, i.e, near-UV and optical; \citealt{Calvet_Gullbring_1998,Bouvier_2007,Herczeg_Hillenbrand_2008}) and line emission along a broad wavelength range (e.g., Balmer series as well as other hydrogen recombination lines such as Pa$\beta$ and Br$\gamma$, and metallic lines as He I and Ca II; \citealt{Alcala_2014}).  
The Paschen continuum emission veils photospheric absorption lines at optical and near-infrared wavelengths. 
Veiling causes the photospheric absorption lines to appear shallower compared to non-accreting stars since the additional emission reduces the contrast between the absorption lines and the continuum \citep{Hartmann_2016}. Recent work by \citet{Robinson_Espaillat_2019}, as well as \citet{Espaillat_2021} or \citet{Pittman_2022}, has shown that the accretion shock could be composed by multiple flows at different densities, with low density components leading to higher veiling in the red part of the spectra (from Y band on), possibly causing the higher veiling observed in classical T Tauri stars (e.g., \citealt{Fischer_2011}). We acknowledge that our modeling does not explicitly account for these multiple density components, but comparison between the method used here and the multi-flow shock modeling suggest a good agreement between both (e.g., \citealt{Pittman_2025}).

To disentangle the effects of extinction and accretion, we use FRAPPE (FitteR for Accretion ProPErties; \citealt{Claes_2024}), a tool that allows us to derive stellar parameters, accretion properties, and extinction from X-Shooter spectra in a self-consistent way. FRAPPE is built upon the recipe presented in \citet{Manara_2013a}. The algorithm uses a grid of non-accreting Class III YSO templates spanning SpTs from G5 to M9.5 compiled from \citet{Manara_2013b}, \citet{Manara_2017_templates}, and \citet{Claes_2024}. A refined sequence of templates between G8 and M9.5 is adopted, based on the implementation of \citet{Claes_2024}, where additional intermediate templates were generated through interpolation to improve SpT coverage within this range. The Class III templates are used to reproduce the photospheric contribution of the host star whilst the modeling of the accretion contribution and the extinction (A$_{\rm V}$) are done using a grid of isothermal hydrogen slab models \citep{Valenti_1993,Rigliaco_2012,Manara_2013a} and the Cardelli extinction law~\citep[R$_{\rm V}$=3.1 \footnote{We acknowledge that the variability of R$_{\rm V}$ across the Orion Nebula is still a matter of debate. Since our sources lie outside the dense core of the Trapezium Cluster, we assume R$_{\rm V}$ = 3.1, in line with previous works \citep[e.g.,][]{DaRio_2010}.};][]{Cardelli_1989}, respectively. This approach of modeling the UV continuum has been widely implemented and shown to effectively capture key characteristics of accretion (e.g., \citealt{Herczeg_Hillenbrand_2008,Rigliaco_2012,Manara_2013b,Manara_2016,Manara_2017_ChaI,Alcala_2014,Alcala_2017,Venuti_2019,Claes_2024}).

In brief, the best fit to each individual observed spectrum is obtained by finding the combination of Spectral Type (SpT), A$_{\rm V}$, and slab parameters (electron temperature, electron density and the optical depth at 300nm) along with two normalisation constants (H$_{\rm slab}$ and K$_{\rm cl3}$) that minimizes a $\chi^{2}_{\rm like}$ distribution. 
These normalisation constants ensure that the templates and slab models are properly scaled to match the observed de$-$reddened spectrum: H$_{\rm slab}$ adjusts the slab emission both for the emitting area (set by the slab parameters) and for the distance to the source, while K$_{\rm cl3}$ rescales the flux of the photospheric template according to the observed target's radius and distance.
The wavelength ranges considered for the fitting encompass features such as the Balmer jump, Balmer and Paschen continua and several TiO bandheads around $\sim$710 nm. We refer to \citet{Claes_2024} for further specific details on the method.

\subsubsection{Spectral type, stellar luminosity and extinction}

Firstly, we estimated the SpTs of our 40 targets by computing the narrow-band TiO spectral index at $\sim$7140 $\AA$ \citep{Oliveira_2003,Jeffries_2007} on the X-Shooter spectra. This spectral index correlates positively for SpTs from late-K ($>$K6) down to M types. At higher temperatures, i.e., earlier SpTs than late-K, TiO is not present in stellar photospheres. The TiO spectral index is computed by calculating the ratio between the pseudo-continuum flux integrated within the range $[$702,705$]$ nm and the strength of the TiO molecular band integrated within the range $[$712.5,715.5$]$ nm. The SpT estimates using this spectral index are shown in Table \ref{tab:fitter_output}. 

We used the SpTs estimated from the TiO index to narrow the grid of Class III templates considered by FRAPPE when fitting each spectrum. We let the SpT free to vary between $\pm$3 sub-SpTs of the previously derived TiO-index SpT. The extinction parameter is free to vary between 0 and 3 mag with a step of 0.1 mag (Orion Nebula members are typically extincted by A$_{\rm V}\lesssim$ 3 mag, \citealt{DaRio_2010,Scandariato_2011}). 
From FRAPPE, we directly obtain the SpT and A$_{\rm V}$ for each source. The typical uncertainties for the fitting procedure are $\sim$0.5 subclasses for the SpT and $\sim$0.2 mag for A$_{\rm V}$ \citep{Claes_2024}. All sources in our sample have extinction values below 3 mag, with the highest estimated A$_{\rm V}$ being 2.8 mag (star \#5), confirming that the chosen grid adequately spans the extinction range of our targets. Out of our sample of 40 PMS stars, 8 have previously reported spectral types based on spectral indices, particularly those tracing TiO and VO absorption bands in the red-optical, from \citet{Hillenbrand_2013}. The spectral types we derive are consistent with theirs to within $\lesssim$0.5 subclasses.

The effective temperature, T$_{\rm eff}$, is determined from the SpT$-$T$_{\rm eff}$ relation from \citet{Herczeg_Hillenbrand_2014} and the bolometric flux, F$_{\rm bol}$, from their bolometric correction, which is based on the stellar flux at 751 nm. For stars with T$_{\rm eff}<$4500 K, FRAPPE adopts a revision to the original calibrations (see \citealt{Claes_2024} for details). The flux at 751 nm used for this correction corresponds to the photospheric emission only. FRAPPE computes it by taking the de-reddened observed flux at 751 nm and subtracting the contribution of the respective best-fit slab model at the same wavelength. The stellar luminosity is then derived from $L_{\star} = 4\pi d^{2} \rm F_{\rm bol}$, where the distance $d$ (in parsecs) is obtained by inverting the \textit{Gaia} parallax of each individual star. Typical uncertainties on the luminosity are around 0.2 dex \citep{Manara_2017_ChaI}. The stellar radius is computed through $R_{\star} = \sqrt{L_{\star}/(4\pi \sigma \rm T_{\rm eff}^{4})}$.

Finally, the stellar masses ($M_{\star}$) and ages were estimated by interpolating each target's position on the Hertzsprung-Russell (HR) diagram adopting the luminosity and effective temperature derived with FRAPPE against a set of PMS theoretical evolutionary models. We considered several sets of models \citep{Dantona_1994, Palla_1999, Siess_2000, PISA_isochrones, Baraffe_2015, Feiden_2016}, and although the absolute values of age for an individual object may slightly vary, we consistently find that the X-Shooter sample is younger than $\lesssim$5 Myr. The only exceptions are three sources (\#30, \#38, and \#39) that appear older than 10 Myr (see Section~\ref{sec:discussion} for further discussion). To be consistent with the CMD presented in Figure \ref{fig:cmd}, hereafter we will adopt the PISA stellar models \citep{PISA_isochrones}. This set of isochrones, properly computed to study the stellar population in Orion \citep[see Section 4.2 of][]{Jerabkova_2019}, provides theoretical magnitudes in the $r$ and $i$ bands of OmegaCAM, ensuring compatibility with the photometric system used in our analysis. While a detailed comparison of different evolutionary models could offer valuable insights, it is beyond the scope of this paper. The stellar masses across the sample span the range $\sim$0.1–0.8 M$_{\odot}$. The final stellar parameters for all targets are summarized in Table~\ref{tab:fitter_output}.

\renewcommand{\arraystretch}{1.2}

\begin{table*}[h!]
\caption{Parameters derived from FRAPPE for the X-Shooter sample.}
\begin{center}
\resizebox{\textwidth}{!}{
\begin{tabular}{lllllllllllll}
\hline \hline
ID & RA & DEC & SpT index & SpT & T$_{\rm eff}$ [K] & A$_{\rm V}$ [mag] & $L_{\star}$ [$L_{\odot}$] & $R_{\star}$ [$R_{\odot}$] & $\log$($L_{\rm acc}$/$L_{\odot}$) & $M_{\star}$ [$M_{\odot}$] & Age [Myr] & $\dot{M}_{\rm acc}$ [$M_{\odot}$/yr] \\ \hline
1   & 05:35:16.1 & $-$05:20:36 & K7  & M0.0 & 3900 & 1.7 & 1.07 & 2.27 & -0.85 & 0.48 & 0.6 & 2.66$\cdot 10^{-8}$ \\
2   & 05:34:52.2 & $-$05:22:32 & M2  & M2.0 & 3560 & 0.8 & 0.87 & 2.45 & -1.55 & 0.32 & 0.4 & 8.55$\cdot 10^{-9}$ \\
3   & 05:35:57.7 & $-$05:57:18 & M0  & M0.5 & 3810 & 0.4 & 0.26 & 1.17 & -1.09 & 0.58 & 4.7 & 6.60$\cdot 10^{-9}$ \\
4   & 05:35:08.0 & $-$05:32:44 & K7  & K6.5 & 4067 & 1.5 & 0.45 & 1.35 & -1.24 & 0.74 & 3.9 & 4.13$\cdot 10^{-9}$ \\
5   & 05:34:26.2 & $-$05:26:30 & K5  & K6.5 & 4067 & 2.8 & 1.34 & 2.33 & -1.21 & 0.57 & 0.6 & 1.00$\cdot 10^{-8}$ \\
6   & 05:35:03.6 & $-$05:29:26 & M1  & M1.5 & 3640 & 0.8 & 0.26 & 1.28 & -1.77 & 0.43 & 2.6 & 2.02$\cdot 10^{-9}$ \\
7   & 05:35:31.5 & $-$05:05:02 & M3  & M3.0 & 3410 & 0.2 & 0.26 & 1.46 & -1.73 & 0.27 & 1.3 & 3.95$\cdot 10^{-9}$ \\
8   & 05:35:25.2 & $-$05:15:36 & M1  & M1.5 & 3640 & 2.1 & 0.61 & 1.96 & -1.25 & 0.39 & 0.8 & 1.13$\cdot 10^{-8}$ \\
9   & 05:35:28.1 & $-$05:29:34 & K5  & ...  & ...  & ... & ...  & ...  & ...   & ...  & ... & ... \\
10  & 05:34:13.2 & $-$05:33:54 & M0  & M1.5 & 3640 & 0.1 & 0.18 & 1.01 & -1.47 & 0.38 & 4.6 & 3.23$\cdot 10^{-9}$ \\
11  & 05:35:32.2 & $-$05:44:27 & M3  & M2.5 & 3485 & 0.5 & 0.46 & 1.86 & -2.54 & 0.29 & 0.8 & 7.40$\cdot 10^{-10}$ \\
12  & 05:34:45.2 & $-$05:10:48 & M1  & M2.0 & 3560 & 1.1 & 0.36 & 1.58 & -1.23 & 0.35 & 1.2 & 1.07$\cdot 10^{-8}$ \\
13  & 05:34:50.5 & $-$05:20:20 & K7  & K6.5 & 4067 & 1.5 & 0.56 & 1.51 & -1.12 & 0.72 & 2.7 & 6.35$\cdot 10^{-9}$ \\
14  & 05:35:41.3 & $-$05:27:50 & M1  & M3.0 & 3410 & 0.6 & 0.16 & 1.15 & -1.29 & 0.28 & 2.3 & 8.39$\cdot 10^{-9}$ \\
15  & 05:35:19.0 & $-$05:21:08 & M3  & M2.5 & 3485 & 1.6 & 0.57 & 2.01 & -2.40 & 0.30 & 0.7 & 1.06$\cdot 10^{-9}$ \\
16  & 05:35:30.5 & $-$05:28:31 & M1  & M1.5 & 3640 & 1.2 & 0.44 & 1.67 & -1.20 & 0.38 & 1.1 & 1.09$\cdot 10^{-8}$ \\
17  & 05:35:29.3 & $-$05:25:46 & M3  & M3.0 & 3410 & 0.4 & 0.54 & 2.11 & -2.45 & 0.26 & 0.5 & 1.12$\cdot 10^{-9}$ \\
18  & 05:35:36.4 & $-$05:31:38 & M2  & M4.0 & 3190 & 1.2 & 0.34 & 1.91 & -1.02 & 0.17 & 0.2 & 4.18$\cdot 10^{-8}$ \\
19  & 05:34:56.5 & $-$05:27:51 & M1  & M1.5 & 3640 & 0.9 & 0.34 & 1.47 & -2.24 & 0.40 & 1.6 & 8.29$\cdot 10^{-10}$ \\
20  & 05:35:22.6 & $-$05:14:11 & M3  & M2.5 & 3485 & 2.2 & 0.85 & 2.53 & -1.47 & 0.29 & 0.2 & 1.16$\cdot 10^{-8}$ \\
21  & 05:35:23.7 & $-$05:26:27 & M1  & M3.5 & 3300 & 2.2 & 0.36 & 1.84 & -0.82 & 0.22 & 0.6 & 5.06$\cdot 10^{-8}$ \\
22  & 05:34:42.7 & $-$05:28:38 & M4  & M3.5 & 3300 & 0.4 & 0.33 & 1.76 & -3.03 & 0.22 & 0.8 & 3.05$\cdot 10^{-10}$ \\
23  & 05:35:17.4 & $-$05:42:15 & M3  & M3.5 & 3300 & 0.6 & 0.27 & 1.59 & -1.80 & 0.21 & 1.0 & 4.86$\cdot 10^{-9}$ \\
24  & 05:34:52.9 & $-$05:28:59 & M2  & M3.0 & 3410 & 0.6 & 0.15 & 1.11 & -1.60 & 0.28 & 2.5 & 4.03$\cdot 10^{-9}$ \\
25  & 05:35:05.8 & $-$05:33:56 & M4  & M4.0 & 3190 & 1.0 & 0.23 & 1.57 & -2.23 & 0.17 & 0.6 & 2.11$\cdot 10^{-9}$ \\
26  & 05:35:24.5 & $-$05:26:32 & M4  & M4.0 & 3190 & 1.4 & 0.42 & 2.12 & -2.22 & 0.15 & 0.1 & 3.30$\cdot 10^{-9}$ \\
27  & 05:35:19.3 & $-$05:16:45 & M4  & M3.5 & 3300 & 1.1 & 0.16 & 1.22 & -2.50 & 0.22 & 1.7 & 7.14$\cdot 10^{-10}$ \\
28  & 05:35:06.0 & $-$05:27:11 & K7  & ...  & ...  & ... & ...  & ...  & ...   & ...  & ... & ... \\
29  & 05:35:06.6 & $-$05:26:51 & M5  & M4.5 & 3085 & 0.6 & 0.30 & 1.92 & -3.03 & 0.13 & 0.1 & 5.67$\cdot 10^{-10}$ \\
30  & 05:35:25.5 & $-$05:45:45 & K7  & K6.5 & 4067 & 1.6 & 0.23 & 0.97 & -1.34 & 0.79 & 14.4 & 2.21$\cdot 10^{-9}$ \\
31  & 05:34:49.1 & $-$05:26:27 & M3  & M4.0 & 3190 & 0.6 & 0.11 & 1.09 & -2.06 & 0.16 & 1.8 & 2.38$\cdot 10^{-9}$ \\
32  & 05:35:06.7 & $-$05:25:03 & M3  & ...  & ...  & ... & ...  & ...  & ...   & ...  & ... & ... \\
33  & 05:34:49.6 & $-$05:05:00 & M4  & M4.0 & 3190 & 0.1 & 0.10 & 1.04 & -3.01 & 0.15 & 1.9 & 2.63$\cdot 10^{-10}$ \\
34  & 05:34:59.1 & $-$05:44:30 & M1  & M1.5 & 3640 & 0.5 & 0.69 & 2.09 & -1.57 & 0.39 & 0.7 & 5.78$\cdot 10^{-9}$ \\
35  & 05:35:14.0 & $-$05:52:09 & K7  & K6.0 & 4115 & 0.2 & 0.61 & 1.54 & -2.11 & 0.75 & 2.6 & 6.42$\cdot 10^{-10}$ \\
36  & 05:35:25.5 & $-$04:51:21 & M2  & M3.5 & 3300 & 1.4 & 0.25 & 1.53 & -1.46 & 0.21 & 1.1 & 1.02$\cdot 10^{-8}$ \\
37  & 05:35:30.9 & $-$04:55:18 & K7  & K7.5 & 3960 & 0.5 & 0.55 & 1.58 & -1.10 & 0.61 & 2.0 & 8.14$\cdot 10^{-9}$ \\
38  & 05:32:49.9 & $-$06:10:46 & M0  & M1.5 & 3640 & 0.5 & 0.05 & 0.56 & -1.74 & 0.47 & 38.1 & 8.64$\cdot 10^{-10}$ \\
39  & 05:33:21.0 & $-$06:22:33 & K7  & M1.5 & 3640 & 1.4 & 0.06 & 0.62 & -1.30 & 0.48 & 29.5 & 2.55$\cdot 10^{-9}$ \\
40  & 05:34:12.9 & $-$05:28:48 & K7  & K6.5 & 4067 & 0.3 & 0.48 & 1.40 & -2.13 & 0.74 & 3.5 & 5.59$\cdot 10^{-10}$ \\
\hline
\end{tabular}}
\end{center}
\label{tab:fitter_output}
\end{table*}

While we assume the inverse of the \textit{Gaia} DR3 parallax as the distance of each star analysed with FRAPPE, we also assessed the impact on the stellar parameter derived if a fixed distance of 390 pc was adopted for all targets in this study. In either case, most of the population shows ages younger than 5 Myr, with only three stars (\#30, \#38, and \#39) appearing significantly older ($>$10 Myr). A few objects in the sample have RUWE values\footnote{The RUWE (Renormalised Unit Weight Error) quantifies the quality of \textit{Gaia}’s astrometric fit; values $\lesssim$1.4 typically indicate well-behaved single-star solutions, while higher values may suggest binarity or other perturbing effects \citep{Lindegren_2021}.} larger than 1.4, suggesting potentially unreliable astrometry. Still the inclusion of these sources is motivated by their infrared excess and the presence of accretion signatures, which justify their relevance to our study.

\subsubsection{Accretion properties}

The accretion luminosity, $L_{\rm acc}$, is obtained by FRAPPE considering the best-fit slab model. The slab models extend below the minimum wavelength of the X-Shooter observations (300 nm). Hence, while the best slab model is found by fitting the model itself against the wavelength coverage of X-Shooter, the total accretion flux, $F_{\rm acc}$, is obtained by integrating over the entire wavelength range provided by the best-fit slab model (50$-$2500 nm). The total accretion flux can be used to obtain $L_{\rm acc}$ through $L_{\rm acc} = 4\pi d^{2} F_{\rm acc}$ (see Table \ref{tab:fitter_output}). The mass accretion rate, $\dot{M}_{\rm acc}$, is computed using equation \ref{eq:macc}: 

\begin{equation} \label{eq:macc}
    \dot{M}_{\rm acc} = \frac{ L_{\rm acc} R_{\star} }{ GM_{\star}} \left( 1 - \frac{R_{\star}}{R_{\rm in}} \right)^{-1} \approx 1.25 \frac{L_{\rm acc}R_{\star}}{G M_{\star}}
\end{equation}

We assume that the YSO inner disc radius, $R_{\rm in}$, is equal to 5R$_{\star}$, following the convention adopted in the magnetospheric accretion model \citep[e.g.,][]{Gullbring_1998} and commonly assumed in subsequent studies (e.g, \citealt{Herczeg_Hillenbrand_2008}; \citealt{Rigliaco_2012}; \citealt{Alcala_2014}; \citealt{Manara_2020}).

We were able to estimate stellar and accretion properties for 37 out of the initial sample of 40 sources using FRAPPE. We show all the best fits of the X-Shooter spectra in the Balmer jump region in Appendix \ref{app:FRAPPE_fits} as well as the spectra of the 3 targets for which we could not find a suitable fit (IDs \#9, \#28, and \#32). In the first source, emission from the accretion process appears to dominate over the photospheric emission, effectively masking the molecular absorption bands (e.g., TiO, CaH and VO) used by FRAPPE for spectral classification (see Figure \ref{fig:all_FRAPPE_fits}).
On the other hand, FRAPPE could not be implemented in the other two sources due to low SNR in the UVB arm (SNR$<$6 at 400 nm whereas all the other sources in the sample have SNR$>$10). Since one of the goals of this work is to carry out a homogeneous analysis of accretion properties based on UV continuum excess fitting, we decided not to include these sources in the following analysis. Unfortunately, the three sources are part of the seemingly old population according to the CMD (see Figure \ref{fig:cmd}), leaving us with 11 sources from this population with estimated stellar and accretion properties from UV excess modeling. 
The accretion luminosities and rates derived for the 37 stars with good fits fall within the following ranges: $-$3.0$\lesssim \log$($L_{\rm acc}$/$L_{\odot})\lesssim-$0.8 and 2$\times$10$^{-10} \lesssim \dot{M}_{\rm acc}$$\lesssim$ 5$\times$10$^{-8}$ M$_{\odot}$/yr. 
The typical uncertainties for $L_{\rm acc}$ and $\dot{M}_{\rm acc}$ are $\sim$0.25 dex and $\sim$0.35 dex, respectively \citep{Claes_2024}. While we will provide in Section \ref{subsec:acc_relations} a detailed discussion of the accretion properties of the stars studied in this work, we anticipate here that the values we find using FRAPPE are in line with typical values found in the literature for PMS stars in the same age and mass range \citep{Manara_2023_PPVII}.

As a consistency check, we also derived L$_{\rm acc}$ from multiple emission lines using empirical L$_{\rm line}-$L$_{\rm acc}$ relations. For most sources, the line-based accretion luminosities are in good agreement with those obtained from UV excess modeling, with a mean difference and spread of 0.15$\pm$0.36 dex. We emphasize that throughout the paper we adopt the L$_{\rm acc}$ values derived from the UV excess as our reference.

\subsection{Lithium feature at $\lambda$670.8 nm}\label{subsec:lithium}

Lithium ($^{7}$Li, hereafter Li) is a short-lived element in the photosphere of low-mass PMS stars since it is depleted when core temperatures reach $\sim$3$\times$10$^{6}$ K \citep{Soderblom_2014}. For this reason, equivalent-widths of the Li line (EW$_{\rm Li}$) at $\lambda$670.8 nm have been widely used for identifying young stars (e.g., \citealt{Jeffries_2023}). Here we compute EW$_{\rm Li}$ for our sample considering that, as explained in Section \ref{sec:FRAPPE}, the spectra of YSOs hosting a protoplanetary disc and undergoing mass accretion, are affected by veiling. This effect needs to be taken into account when measuring reliable EW$_{\rm Li}$.

Using the best-fit slab models obtained from FRAPPE, we perform veiling correction on the X-Shooter spectra. First, each spectrum is de-reddened using the best-fit value of A$_{\rm V}$. Next, the wavelength step of the best-fit slab model is resampled to match that of the observed spectrum. The resampled best-fit slab model is scaled to the observed spectrum by multiplying it by the normalisation constant H$_{\rm slab}$. Finally, the rescaled best-fit slab model is subtracted from the de-reddened observed spectrum, resulting in the veiling-corrected spectrum.
We compute the EW$_{\rm Li}$ by first normalizing the pseudo-continuum around the spectral feature using a second-degree polynomial fit in the region [670.4, 671.2] nm, excluding the lithium line itself from the fit. The standard deviation of the continuum ($\sigma_{\rm spec}$) is taken as the noise of the spectrum. To estimate the uncertainty, we apply a Monte Carlo approach where the flux is perturbed by adding values sampled from a Gaussian distribution (with $\mu$=0 and $\sigma$=$\sigma_{\rm spec}$) in each iteration. The equivalent width is calculated by integrating the line after each perturbation. This process is repeated multiple times, and the final EW$_{\rm Li}$ is taken as the mean of the measurements, with its uncertainty as the standard deviation.

We detect lithium absorption in the majority of our sample except in two sources, \#9 and \#32 (see Figure \ref{fig:lithium_sample}). Source \#9 appears to be a strong accretor, and the non-detection of the Li I line may be due to strong veiling, which can significantly weaken photospheric absorption features.  
As for \#32, we could not detect Li I in its spectrum due to low SNR. For the remaining sources the presence of lithium supports their young nature.

\begin{figure}[ht!]
\centering
    \includegraphics[width=0.5\textwidth]{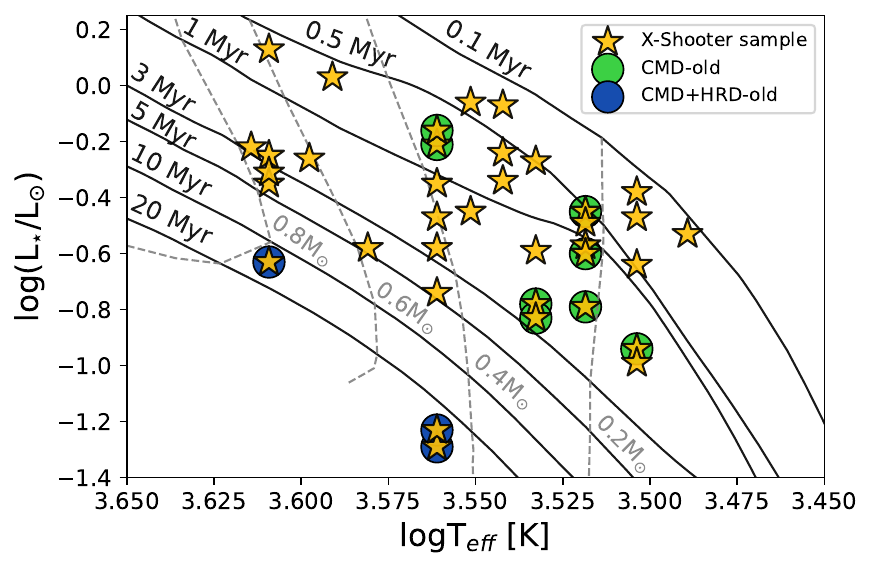}
\caption{HR diagram of our sample (yellow stars) with sources appearing older than 10 Myr in the CMD (CMD$-$old; see Figure \ref{fig:cmd}) highlighted with green circles. Among these sources, those remaining older than 10 Myr after estimating stellar parameters with FRAPPE (CMD$+$HRD$-$old) are marked with dark blue circles. The 0.1, 0.5, 1, 3, 5, 10 and 20 Myr PISA isochrones \citep{PISA_isochrones} are shown as solid black lines, while the 0.2, 0.4, 0.6 and 0.8 M$_{\odot}$ mass tracks are presented as dashed gray lines.}
\label{fig:HRD}
\end{figure}

\section{Discussion}\label{sec:discussion}

In this section, we analyse the isochronal ages obtained through the position of the sources in the HR diagram, based on the newly derived stellar parameters using FRAPPE on the X-Shooter spectra. Additionally, the equivalent-width measurements of the lithium absorption feature are used to further corroborate the young nature of our sample. We also explore relationships between accretion properties and host star characteristics by comparing our derived parameters with those measured in other nearby SFRs. Finally, we inspect the impact of accretion on the optical colours of our sample.

\subsection{Youth indicators: HR diagram and lithium}

In Figure \ref{fig:HRD}, we show the distribution of the stars on the HR diagram (HRD). We adopt the effective temperatures and luminosities obtained using the multicomponent fitting procedure FRAPPE (see Section \ref{sec:FRAPPE}). PISA isochrones and evolutionary tracks \citep{PISA_isochrones} are included for reference. Most stars in our sample have estimated ages younger than 3 Myr, consistent with the expected age of the Orion Nebula population, with only a few falling in the 3$-$5 Myr range. However, there are three sources in our sample (IDs \#30, \#38, and \#39) that stand out by appearing significantly older in the HRD ($\gtrsim$10 Myr; highlighted as dark blue solid circles in Figure~\ref{fig:HRD}). We refer to this subgroup as the "CMD+HRD$-$old" subset in the remainder of the analysis.

Unlike the rest of the sample, where a self-consistent treatment of extinction and accretion reconciles the stars with younger estimated ages, sources \#30, \#38, and \#39 remain significantly older despite the application of the same fitting procedure. All three stars show clear signatures of disc presence and ongoing mass accretion, suggesting that they are not evolved field contaminants. Moreover, although these sources are located toward the southern outskirts of the surveyed region (at declinations below $-6^\circ00'$; see Figure~\ref{fig:ON_fov}), their proper motions are fully consistent with those of the main population, and their RUWE values are within the expected range for reliable astrometric solutions.

Even excluding the three most underluminous sources, we still observe a spread of luminosity in the HRD shown in Figure \ref{fig:HRD}. This spread reaches up to $\sim$0.7 dex if measured as the full vertical extent in L$_{\rm \star}$ at a given T$_{\rm eff}$. A spread in stellar luminosity within Orion Nebula stars has been reported by several previous works \citep[e.g.][]{DaRio_2010,Jeffries_2011}. Different processes can contribute to this spread, including a complex star formation history, variations in accretion activity of YSOs, disc geometry with respect to the observer and star spots \citep[e.g.][]{Beccari_2017,Jerabkova_2019,baraffe_2017,Guarcello_2010,Franciosini_2022}.

\begin{figure}[h]
    \includegraphics[width=\columnwidth]{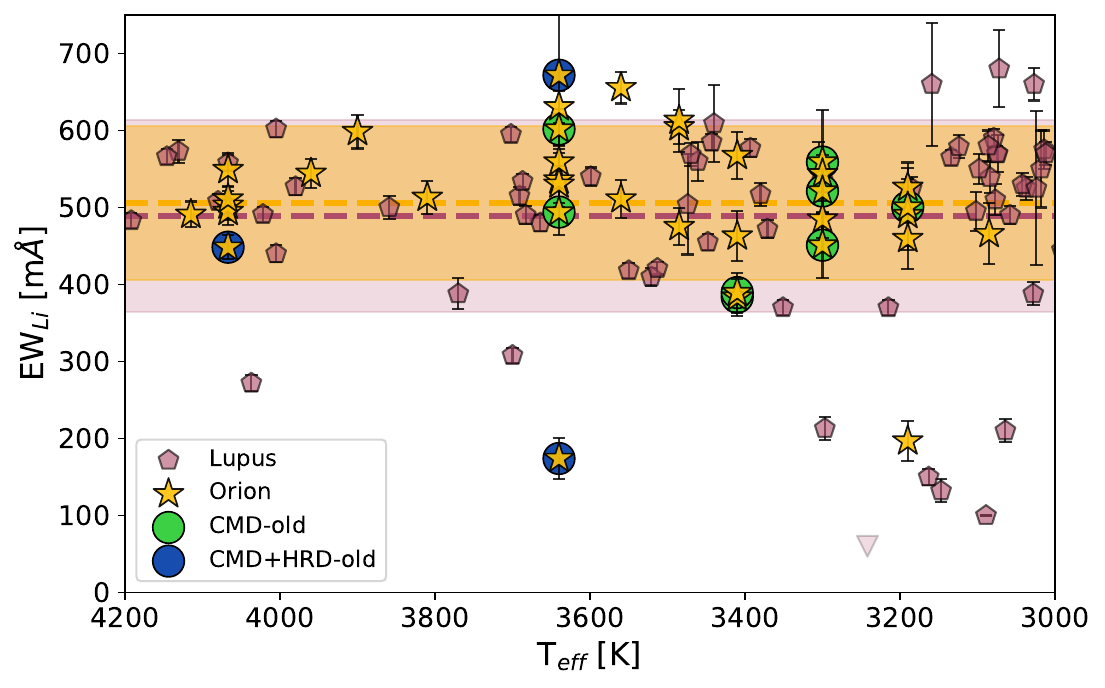}
    \caption{EW$_{\rm Li}$ vs. T$_{\rm eff}$ for the veiling-corrected sample as obtained with FRAPPE. The marker scheme is the same as in Figure \ref{fig:HRD}. The pink pentagons are EW$_{\rm Li}$ measurements performed on X-Shooter spectra of Class II sources located in Lupus from \citet{Biazzo_2017}. The mean EW$_{\rm Li}$ values and associated standard deviations in the T$_{\rm eff}$ range between 4200 K down to 3000 K are plotted in yellow and pink for the Orion Nebula and Lupus, respectively.}
    \label{fig:ewLi_teff}
\end{figure}

To further investigate the age distribution of our sample, we computed veiling-corrected EW$_{\rm Li}$, a common youth indicator in PMS stars. We show in Figure \ref{fig:ewLi_teff} the EW$_{\rm Li}$ vs. T$_{\rm eff}$ measured in this work using X-Shooter spectra for the stars in Orion, together with measurements from sources located in the Lupus star-forming complex. Lupus, located within 200 pc of the Sun, is one of the prominent low-mass SFRs and hosts stars with estimated ages between 1 and 3 Myr \citep[see review by][]{Comeron_2008}. Given its similar age range to the Orion Nebula, Lupus provides an ideal benchmark for evaluating the overall lithium depletion in our sample. The EW$_{\rm Li}$ measurements for Lupus stars were obtained from X-Shooter spectra by \citet{Biazzo_2017}, based on a dataset of Class II and Class III targets presented in previous works \citep{Alcala_2014,Alcala_2017,Frasca_2017}. For consistency, we include only Class II Lupus sources in our comparison with the sample of \citet{Biazzo_2017}.

In the same figure, we show the mean and standard deviation of the distribution of the EW$_{\rm Li}$ in the two samples for stars in the temperature range 3000$< \rm T_{\rm eff} <$ 4200 K. For the Orion Nebula, we find $<$EW$_{\rm Li}>=505 \pm 100$ m$\AA$ in agreement with previous studies \citep[see e.g.][]{Palla_2005}, while for Lupus, we find $<$EW$_{\rm Li}>=489 \pm 125$ m$\AA$. The properties of the EW$_{\rm Li}$ of the populations in the two SFRs are statistically consistent, suggesting a similarity in the median age of their stellar populations.

However, two Orion stars stand out for having weaker Li lines compared to the rest of the population: \#18 with EW$_{\rm Li} = 196 \pm 26$ m$\AA$, and \#38 with EW$_{\rm Li} = 174 \pm 27$ m$\AA$. Source \#38 (T$_{\rm eff}\sim$3640 K) is among the oldest stars in the HRD and, together with \#39, is one of the two most underluminous sources in our sample (see Figure \ref{fig:HRD}). Despite having the same estimated effective temperature, source \#38 shows significant lithium depletion, whereas source \#39 displays the strongest lithium absorption among all stars with similar T$_{\rm eff}$. On the other hand, source \#18 (T$_{\rm eff}$ $\sim$3190 K) appears extremely young in the HRD, with an isochronal age of only 0.2 Myr, and low mass ($\sim$0.2 M$_{\odot}$), making the low EW$_{\rm Li}$ more difficult to interpret.

These examples highlight the need for caution when using lithium equivalent width as a standalone age indicator for individual stars at very young ages ($<$10 Myr), since no significant depletion is expected at this stage. In this context, EW$_{\rm Li}$ is more useful as a relative age indicator within a given population, rather than for absolute age determinations. The reliability of lithium depletion as an age tracer for individual stars increases when depletion becomes substantial, typically in PMS stars aged between 20 and 200 Myr \citep{Soderblom_2014}. Given the age range considered here, EW$_{\rm Li}$ cannot be used to derive absolute individual ages, but it provides a valuable comparative tool to assess our Orion Nebula sample's youth relative to other SFRs, such as Lupus. 

\subsection{Accretion relations with host star properties}\label{subsec:acc_relations}

\begin{figure*}[h!]
\centering
    \includegraphics[width=0.5\textwidth]{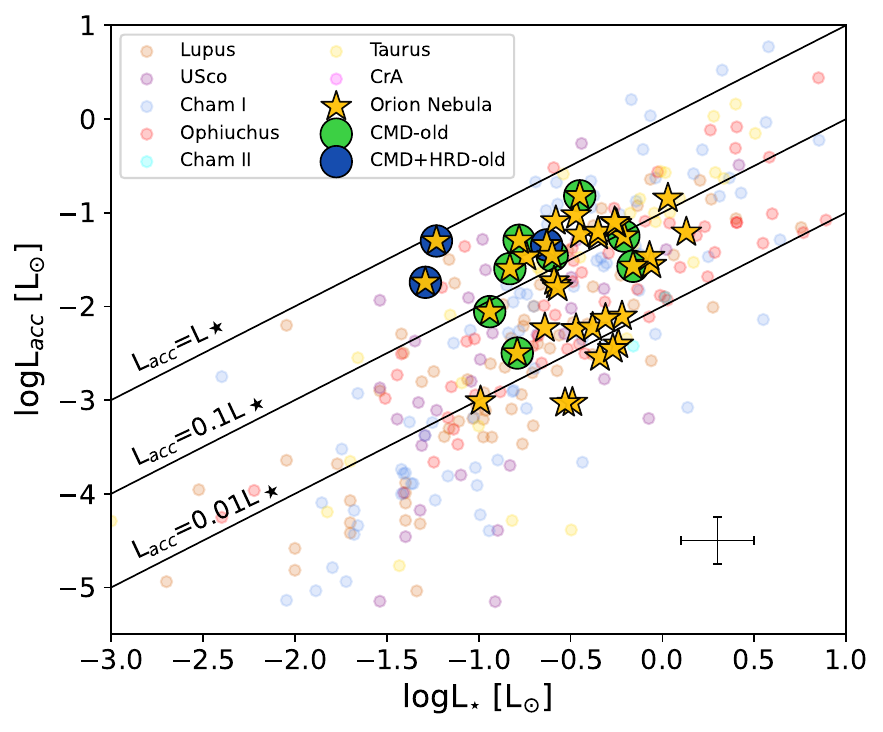}
    \hspace{-15pt}
    \includegraphics[width=0.5\textwidth]{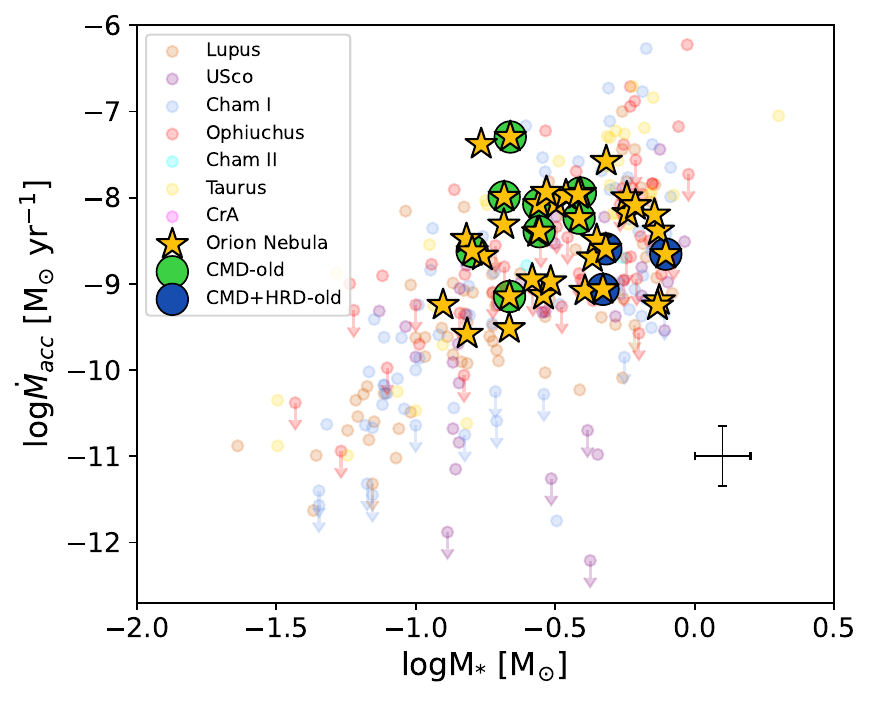}
\caption{Comparison of accretion parameters as a function of stellar properties between our X-Shooter sample of Orion Nebula sources and the sample of different nearby SFRs (transparent circles) presented in \citet{Manara_2023_PPVII}. The marker scheme for our sample is the same as in Figure \ref{fig:HRD}. \textit{Left panel: }Accretion luminosity as a function of stellar luminosity as obtained using FRAPPE. The black lines indicate $L_{\rm acc}$ equal to 1$L_{\star}$, 0.1$L_{\star}$ and 0.01$L_{\star}$. The error bars indicate the typical uncertainties on the stellar and accretion luminosities ($\sim$0.2 dex and 0.25 dex, respectively). \textit{Right panel:} Mass accretion rate as a function of stellar mass as obtained using FRAPPE. Literature targets with arrows pinpoint upper limits. The error bars indicate the typical uncertainties on the mass accretion rate and stellar mass ($\sim$0.35 dex and 0.1 dex, respectively).}
\label{fig:acc_stellar}
\end{figure*}

Previous spectroscopic studies of accreting PMS stars in other nearby SFRs such as Lupus, Chamaeleon I, and Upper Scorpius have extensively characterised accretion properties in lower-mass and less complex (i.e. lower stellar density and FUV radiation from massive stars) environments than that of the Orion Nebula. These studies have revealed a well-defined correlation between accretion and stellar properties across a wide mass range, from intermediate-mass Herbig stars down to the substellar regime (e.g, \citealt{Hillenbrand_1992,Natta_2004,Mohanty_2005,Natta_2006,Alcala_2014,Alcala_2017,Manara_2017_ChaI,Venuti_2019,Alemendros-Abad_2024}). In particular, the scaling of $\dot{M}_{\rm acc}$ with $M_{\star}$ in the logarithmic scale follows a steeper-than-linear relation, with reported slopes in the range of $\sim$1.6$-$2. However, this trend exhibits a significant spread in accretion rates -- up to 1$-$2 dex -- which remains a topic of discussion and has important implications for disc evolution (\citealt{Somigliana_2022,Manara_2023_PPVII} and references therein). While these findings have provided crucial insights into star-disc interactions and the evolution of YSOs, more massive and dynamically complex environments, such as the Orion Nebula, remain comparatively less explored. Conducting similar homogeneous accretion studies in regions like Orion is therefore essential to understand whether the trends observed in lower-mass environments hold in more representative star-forming conditions \citep{Winter_2022} and to further probe the role of accretion in shaping inferred stellar properties.

We show in the left panel of Figure \ref{fig:acc_stellar} the distribution of the accretion luminosity, log$L_{\rm acc}$, as a function of the luminosity of the host star, log$L_{\star}$, for the stars analysed in this work. For comparison, we also include results from previous studies of YSOs in nearby SFRs, as compiled in Table 1 of \citet{Manara_2023_PPVII}. We observe a general trend in which $L_{\rm acc}$ increases with $L_{\star}$. Moreover, all stars lie below the $L_{\rm acc}=L_{\star}$ boundary, indicating that the stellar luminosity exceeds the accretion luminosity across the entire sample. This is the expected behavior for Class II YSOs, where accretion contributes moderately to the total luminosity output. Although higher ratios are more commonly associated with earlier, more embedded evolutionary stages, the majority of observed sources still show accretion luminosities smaller than the stellar luminosity \citep{Clarke_2006,Tilling_2008,Fiorellino_2021}.

We show in the right panel of Figure \ref{fig:acc_stellar} the distribution of $\dot{M}_{\rm acc}$ as a function of M$_{\star}$. The distribution confirms the well-established trend that $\dot{M}_{\rm acc}$ increases with $M_{\star}$. The Orion Nebula sample appears to populate the same parameter space as the one covered by the other young, less dense SFRs, within the typical uncertainties.

The two most underluminous sources in our sample (\#38 and \#39, see Figure \ref{fig:HRD}) show high accretion luminosities relative to their stellar luminosities (log$L_{\star}\sim-$1.3$L_{\odot}$; see upper panel of Figure \ref{fig:acc_stellar}). However, in the $\dot{M}_{\rm acc}$ vs. $M_{\star}$ diagram (bottom panel of the same figure), they appear to have relatively low mass accretion rates for their stellar mass (M$_{\star}\sim$0.4 M$_{\odot}$). Both stars have small estimated radii ($\sim$0.55--0.65 $R_{\odot}$; see Table \ref{tab:fitter_output}). In FRAPPE, the stellar radius is derived from the stellar luminosity following the relation $R_{\star}\propto \sqrt{L_{\star}}$. Thus, if $L_{\star}$ is underestimated, this would lead to an underestimated $R_{\star}$ and, consequently, an underestimated $\dot{M}_{\rm acc}$ (see Equation \ref{eq:macc}). 
While we cannot rule out the possibility that these sources are genuinely old stars with long-lived accretion, an alternative explanation is that their stellar luminosities are underestimated due to significant disc inclination (e.g., \citealt{Alcala_2014}). Nearly edge-on discs can obscure much of the stellar photosphere, reducing the observed luminosity. However, accretion signatures like H$\alpha$ emission and, in some cases, UV excess emission, may remain detectable through scattered light, depending on the disc geometry and accretion rate. Such systems, when observed predominantly in scattered light, can also exhibit anomalously blue colours in CMDs \citep[e.g.,][]{Guarcello_2010, DeMarchi_2013} consistent with what we observe for these two sources.

We conclude by highlighting that all the other sources in our sample that appear old in the CMD $-$ and that have been reassessed as young PMS stars after the analysis presented in this work $-$ demonstrate typical accretion behavior, consistent with their evolutionary stage.

\subsection{Impact of accretion in OmegaCAM photometry}\label{subsec:main_result}

As shown in Figure \ref{fig:HRD}, most PMS stars that appear old in the CMD have estimated stellar parameters that place them in a position on the HRD consistent with a young population ($\sim$1$-$5 Myr). The young age is also supported by strong lithium absorption and typical $L_{\rm acc}$/$L_{\star}$ ratios expected for young, accreting stars. Hence, with our work, we show that the majority of the seemingly old accretors analysed are not intrinsically old, but that their optical magnitudes are strongly affected by the presence of strong accretion luminosity, making these stars appear to have bluer optical colours in the $r, r-i$ CMD (see Figure \ref{fig:cmd}). In the following we describe a method to accurately investigate the impact of accretion on the photometry of our targets by isolating the effect of the accretion process on the light collected through the OmegaCAM $r$ and $i$ filters.
Since OmegaCAM observations were not obtained simultaneously with the X-Shooter data, we adopt the spectroscopic observations as a consistent reference.

We first estimate synthetic photometry by convolving the de-reddened X-Shooter spectra with the response curves of the OmegaCAM $r$ and $i$ filters.  The de-reddening was performed using the best-fit A$_{\rm V}$ values obtained from FRAPPE and applying Cardelli's extinction law. We computed the total flux observed through the $r$ and $i$ broad-bands using the specific filter's throughput and adopting Equation \ref{eq:syn_phot}:

\begin{equation}\label{eq:syn_phot}
    f_{\lambda} [erg s^{-1} cm^{-2} \mathring{A}^{-1}] = \frac{\int (spec \times filter \times \lambda) d\lambda}{\int (filter \times \lambda) d\lambda} 
\end{equation}

where $spec$ is the de-reddened X-Shooter spectrum resampled to the wavelength step of the filter's response curve, $filter$ is the filter's response curve and $\lambda$ the respective wavelength. Once the total flux observed through the $r$ and $i$ band filters is known, one can convert it to magnitudes using $m_{\rm AB}=-$2.5 log $\left( f_{\rm \lambda} / f_{\rm ZP} \right)$ where $f_{\rm r, ZP}=2.7549\times10^{-9}$ erg s$^{-1}$ cm$^{-2}$ $\mathring{A}^{-1}$ and $f_{\rm i, ZP}=1.35102\times10^{-9}$ erg s$^{-1}$ cm$^{-2}$ $\mathring{A}^{-1}$. The filter's response curves and $f_{\rm \lambda, ZP}$ were taken from the Spanish Virtual Observatory (SVO) Filter Profile Service \citep{SVO_service}. 

\begin{figure}[h!]
\centering
    \includegraphics[width=0.5\textwidth]{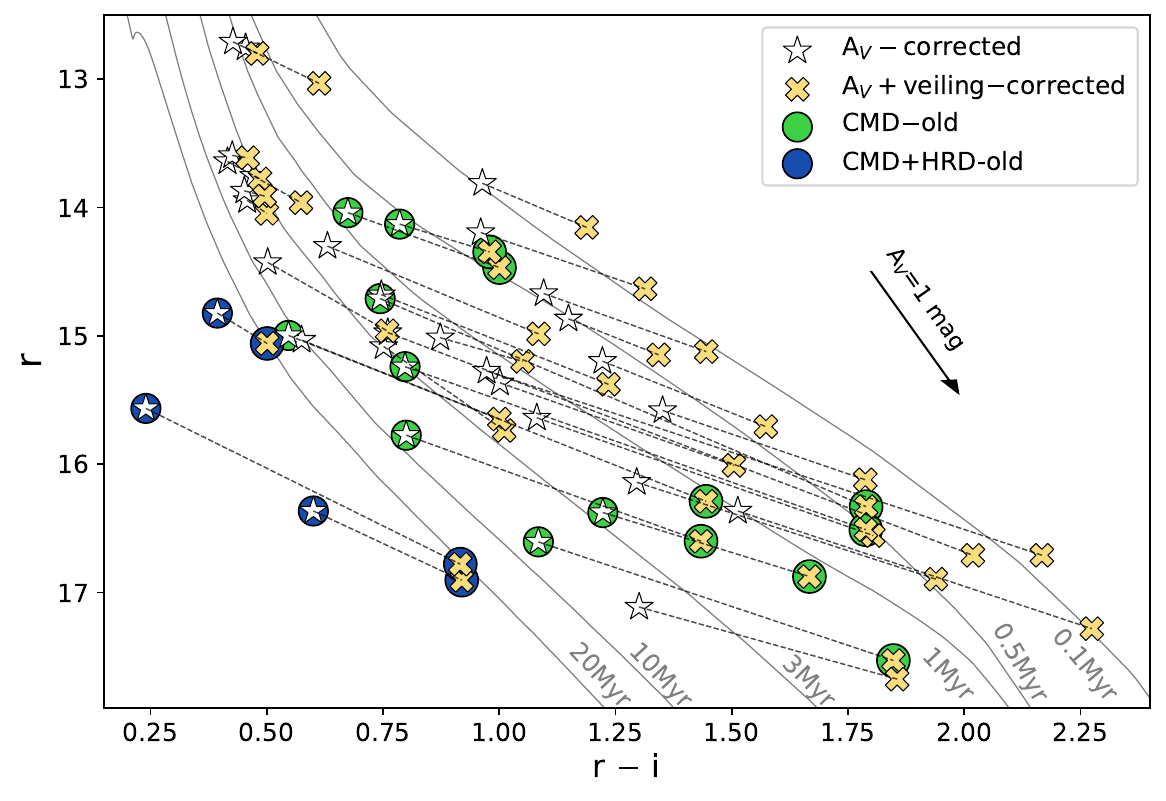}
\caption{$r$ vs. $(r-i)$ CMD illustrating the effect of accretion on the OmegaCAM photometry of our targets. For each star, the white star-like points are derived from synthetic photometry performed on the de-reddened X-Shooter spectra, and include both photospheric and accretion-related emission; the yellow crosses correspond to the expected photospheric emission, estimated by interpolating their positions on the HRD using the stellar parameters derived with FRAPPE and the PISA isochrones. The two measures for each star are connected through a dashed line. We use the same marker scheme as in Figure \ref{fig:HRD} for the CMD$-$old and the CMD$+$HRD$-$ old sub-samples. The black arrow indicates the effect that 1 mag of extinction would have. The solid gray lines show the location of 0.1, 0.5, 1, 3, 10, and 20 Myr PISA isochrones \citep{PISA_isochrones}.}
\label{fig:cmd_new}
\end{figure}

We then estimate the $r$ and $i$ magnitudes of the pure photosphere by performing a weighted interpolation on the HRD, using the T$_{\rm eff}$ and $L_{\star}$ values derived with FRAPPE and the PISA PMS isochrones. This interpolation used a two-dimensional Gaussian kernel centered on the HRD position of each source. These HRD-based magnitudes represent the intrinsic photospheric emission, free from both extinction and, most importantly, accretion contributions, as both effects are accounted for in FRAPPE. By placing both magnitude estimates (i.e., photosphere$+$accretion and photosphere-only) in a $r$ vs. $(r-i)$ CMD and connecting them with a line, we can isolate and visualise the impact of accretion on the OmegaCAM photometry of each star. These results are presented in Figure \ref{fig:cmd_new}.

As shown in the figure, removing the continuum excess emission coming from accretion from the de-reddened spectra (white stars to yellow
crosses) causes most objects to appear fainter and redder. This can be explained by the addition of flux primarily in the bluer part of the spectrum due to the excess emission originating in the accretion shock regions. This figure clearly demonstrates that the presence of accretion luminosity causes a displacement of stars in the optical CMD making accreting PMS stars appear bluer and brighter with respect to their purely photospheric colours. We also show the extinction vector, illustrating that the displacement of stars in these bands due to extinction follows a different direction than the displacement caused by accretion. Hence, both effects must be taken into account when analysing optical CMDs.

We compare the OmegaCAM photometry with the synthetic photometry derived from the X-Shooter spectra, considering only sources for which stellar parameters could be determined with FRAPPE. The two sets of measurements show a clear correlation, though with some scatter and systematic offset. In the $r$ band, we find an absolute mean difference of 0.06 mag with a scatter of 0.43 mag, whereas in the $i$ band, the mean difference is 0.003 mag with a scatter of 0.36 mag. These small offsets indicate the presence of low residuals in the flux calibration of the spectra with respect to the ADHOC catalog while the scatters likely reflect the presence of intrinsic stellar and accretion-related variability. Accreting PMS stars can vary photometrically due to fluctuations in the accretion rate, variable circumstellar extinction, and rotational modulation by stellar spots (e.g., \citealt{Venuti_2015}). As such, some level of discrepancy between photometry taken at different epochs is anticipated.

To assess the impact of accretion on the stellar parameters one might retrieve from the CMD, we applied the same weighted interpolation method as explained above but now based on the PISA isochrones in the CMD to the extinction-corrected magnitudes only. By comparing the stellar parameters obtained accounting and not accounting for accretion, we find mean differences of $\Delta$L$_{\rm \star}\sim$0.32$\pm$0.29 L$_{\rm \odot}$ and $\Delta$T$_{\rm eff}\sim$479$\pm$175 K, which translate into a mean difference of $\sim$0.3 M$_{\odot}$ in M$_{\rm \star}$. These shifts are not restricted to the seemingly older accretors, indicating that accretion affects inferred stellar parameters throughout the sample. Photometric and accretion variability are not disentangled in this analysis. Instead, the accretion-corrected parameters that we report are derived from the X-Shooter spectra alone, which provide a consistent snapshot of each star's accretion and photospheric properties at a single epoch. Veiling, which is a result of the continuum excess, is implicitly accounted for in the spectral modeling and is not treated as a separate parameter.

\section{Summary and Conclusions}

In this work, we present a detailed analysis of VLT/X-Shooter spectra of 40 disc-bearing (according to \textit{Spitzer}) PMS stars located in the Orion Nebula. From these, 14 sources show both isochronal ages older than 10 Myr (with some even $>$30 Myr) in the $r, (r-i)$ CMD, and H$\alpha$ excess emission (from OmegaCAM observations). We obtained accurate stellar and accretion properties of the YSOs in the X-Shooter sample by implementing the multicomponent fitting procedure FRAPPE \citep{Claes_2024} which models the accretion and photospheric contributions using a grid of slab models and a set of interpolated Class III YSO templates, respectively. We were able to fit 37 out of the initial sample of 40 sources. By analysing the parameters retrieved, these are the main results of our study:

\begin{itemize}
    \item The accretion properties of the Orion Nebula PMS stars are consistent with the ones observed in lower-density SFRs such as Lupus, Taurus and Upper Scorpius, both in terms of $L_{\rm acc}$ vs. $L_{\star}$ and $\dot{M}_{\rm acc}$ vs. $M_{\star}$.

    \item After accounting for extinction and accretion, and deriving accurate stellar parameters, the location of most of the seemingly old stars ($\gtrsim10$ Myr) in the HRD is consistent with younger ages ($\sim$1$-$5 Myr). This is further supported by strong lithium features (EW$_{\rm Li} \gtrsim$ 400 m$\AA$), and by typical $L_{\rm acc}/L_{\star}$ ratios expected for young, accreting stars. 

    \item While most sources are consistent with typical ages expected for the Orion Nebula population, three sources (IDs \#30, \#38, and \#39) still hold their old nature ($>$10 Myr) even after our analysis. Additional observations and further analysis will be crucial to determine whether these accreting PMS stars are intrinsically older or their apparent age can be explained by geometric or environmental factors. 

    \item By combining synthetic photometry on the de-reddened X-Shooter spectra with magnitudes interpolated from the HRD (representing the intrinsic photospheric emission), we demonstrated that continuum excess emission from the accretion process significantly alters the observed optical colours of accreting PMS stars. In particular, the $r$ magnitude of stars undergoing accretion appears brighter and their $r-i$ colour is systematically bluer with respect to the pure photospheric colour, virtually making the stars look hotter and hence younger. Such impacts have consequences on the derived stellar luminosity and hence on the stellar masses and ages retrieved.

\end{itemize}

Our work presents a homogeneous study of accretion in PMS stars in the Orion Nebula. The accretion properties we find are in agreement with those found in similarly young populations in low-mass SFRs like Lupus and Taurus. However, a more extensive study covering different stellar and interstellar densities in Orion is needed to better characterise how the environment influences accretion. Moreover, we report how accretion can affect the optical photometry of PMS stars. This pilot study seems to suggest that any investigation of YSOs in SFRs that selects members purely based on their positions in an optical CMD is likely to miss a substantial fraction of objects whose locations fall outside the canonical PMS locus due to strong accretion. Hence, the effect of accretion must be carefully taken into account when estimating the stellar properties from isochrone fitting in such a context. In future work, we will study the impact of accretion across different photometric bands.

\section{Data availability}

Additional data for this article are available in Zenodo at \url{https://zenodo.org/records/17093047}.

\begin{acknowledgements}
We thank the anonymous referee for valuable feedback that improved this manuscript. 
We thank Victor Almendros-Abad for many helpful discussions on these results. 
LP acknowledges the PhD fellowship of the International Max-Planck-Research School (IMPRS) funded by ESO. 

CFM is funded by the European Union (ERC, WANDA, 101039452). Views and opinions expressed are however those of the author(s) only and do not necessarily reflect those of the European Union or the European Research Council Executive Agency. Neither the European Union nor the granting authority can be held responsible for them.

TJ acknowledges the support from the MUNI Award in Science and Humanities.

Based on observations collected at the European Southern Observatory under ESO programme(s) 108.2206.001 and 114.276M.001.

This work has made use of data from the European Space Agency (ESA) mission
{\it Gaia} (\url{https://www.cosmos.esa.int/gaia}), processed by the {\it Gaia} Data Processing and Analysis Consortium (DPAC, \url{https://www.cosmos.esa.int/web/gaia/dpac/consortium}). Funding for the DPAC has been provided by national institutions, in particular the institutions participating in the {\it Gaia} Multilateral Agreement.

This research has made use of the SVO Filter Profile Service "Carlos Rodrigo", funded by MCIN/AEI/10.13039/501100011033/ through grant PID2023-146210NB-I00

This research was supported by the Excellence Cluster ORIGINS, funded by the Deutsche Forschungsgemeinschaft (DFG, German Research Foundation) under Germany’s Excellence Strategy – EXC-2094 – 390783311. We acknowledge the support of the Deutsche Forschungsgemeinschaft (DFG, German Research Foundation) Research Unit “Transition discs” - 325594231.

\end{acknowledgements}

\bibliographystyle{aa}
\bibliography{XShoo_spec}

\begin{appendix} 

\onecolumn

\section{Additional tables}

In Table \ref{tab:omegacam_info} we show the OmegaCAM photometry for the entire X-Shooter sample as obtained within the scope of the Accretion Discs in H$\alpha$ with OmegaCAM (ADHOC) survey \citep{Beccari_2017}.

Table~\ref{tab:source_info} provides the IDs used in this work along with \textit{Gaia} DR3 and OmegaCAM IDs, coordinates, parallax, proper motions, and RUWE values for the sources presented in Section~\ref{sec:data}.

\begin{table}[h]
\caption{OmegaCAM photometry from the ADHOC survey.}
\begin{center}
\resizebox{0.78\textwidth}{!}{
\begin{tabular}{lccccccc}
\hline \hline
ID & OmegaCAM & r\_mag & r\_mag\_err & i\_mag & i\_mag\_err & H$\alpha$\_mag & H$\alpha$\_mag\_err  \\
\hline
1 & 1042831 & 14.25 & 0.01 & 13.42 & 0.01 & 12.8 & 0.01  \\
2 & 1021191 & 14.47 & 0.01 & 13.34 & 0.01 & 13.7 & 0.01 \\
3 & 1079659 & 14.64 & 0.01 & 13.99 & 0.01 & 13.28 & 0.01 \\
4 & 1035737 & 14.76 & 0.01 & 14.2 & 0.01 & 13.76 & 0.01 \\
5 & 996321 & 14.76 & 0.01 & 13.9 & 0.01 & 13.43 & 0.01 \\
6 & 1031924 & 14.85 & 0.01 & 13.96 & 0.01 & 13.71 & 0.01 \\
7 & 1056861 & 14.97 & 0.02 & 14.12 & 0.02 & 13.48 & 0.01 \\
8* & 1051021 & 15.06 & 0.02 & 14.71 & 0.01 & 14.05 & 0.01 \\
9* & 1053477 & 15.72 & 0.01 & 15.25 & 0.01 & 13.88 & 0.01 \\
10 & 985345 & 15.37 & 0.01 & 14.66 & 0.01 & 13.96 & 0.01 \\
11 & 1057420 & 15.46 & 0.01 & 14.1 & 0.01 & 14.73 & 0.01 \\
12 & 1014551 & 15.53 & 0.01 & 14.7 & 0.01 & 13.99 & 0.01 \\
13 & 1019536 & 14.95 & 0.01 & 14.24 & 0.01 & 13.91 & 0.01 \\
14* & 1065824 & 15.56 & 0.01 & 14.97 & 0.01 & 13.85 & 0.01 \\
15 & 1045603 & 15.64 & 0.03 & 14.61 & 0.01 & 11.36 & 0.08 \\
16 & 1055893 & 15.64 & 0.01 & 14.45 & 0.01 & 13.77 & 0.01 \\
17 & 1054793 & 15.66 & 0.01 & 14.26 & 0.01 & 14.97 & 0.03 \\
18 & 1061070 & 15.75 & 0.01 & 14.92 & 0.01 & 14.14 & 0.01 \\
19 & 1025573 & 15.84 & 0.01 & 14.78 & 0.01 & 15.12 & 0.01 \\
20 & 1048863 & 15.84 & 0.01 & 14.51 & 0.01 & 14.13 & 0.01 \\
21* & 1049701 & 15.9 & 0.01 & 15.16 & 0.01 & 14.14 & 0.01 \\
22 & 1011929 & 15.93 & 0.01 & 14.49 & 0.01 & 15.1 & 0.01 \\
23 & 1044085 & 15.97 & 0.01 & 15.0 & 0.01 & 14.49 & 0.01 \\
24* & 1022028 & 16.08 & 0.01 & 15.28 & 0.01 & 14.67 & 0.01 \\
25 & 1033812 & 16.32 & 0.01 & 14.76 & 0.01 & 15.29 & 0.01 \\
26 & 1050335 & 16.34 & 0.01 & 14.97 & 0.01 & 14.54 & 0.01 \\
27* & 1045919 & 16.36 & 0.01 & 15.49 & 0.01 & 14.44 & 0.01 \\
28* & 1033893 & 16.45 & 0.01 & 15.9 & 0.01 & 14.25 & 0.01 \\
29 & 1034451 & 16.5 & 0.01 & 15.04 & 0.01 & 14.96 & 0.02 \\
30** & 1051294 & 16.77 & 0.01 & 16.23 & 0.01 & 15.52 & 0.01 \\
31* & 1018190 & 16.84 & 0.01 & 15.8 & 0.01 & 15.18 & 0.01 \\
32* & 1034406 & 17.04 & 0.04 & 16.56 & 0.01 & 14.68 & 0.05 \\
33 & 1018691 & 17.19 & 0.02 & 15.87 & 0.02 & 15.42 & 0.01 \\
34* & 1028024 & 15.06 & 0.01 & 14.51 & 0.01 & 13.77 & 0.01 \\
35 & 1040907 & 13.86 & 0.01 & 13.34 & 0.01 & 13.18 & 0.01 \\
36* & 1051303 & 15.71 & 0.01 & 15.08 & 0.01 & 13.72 & 0.01 \\
37 & 1056285 & 14.02 & 0.01 & 13.52 & 0.01 & 12.95 & 0.01 \\
38** & 930757 & 16.92 & 0.01 & 16.13 & 0.01 & 15.48 & 0.01 \\
39** & 947152 & 16.13 & 0.01 & 15.44 & 0.01 & 14.64 & 0.01 \\
40 & 985078 & 14.23 & 0.01 & 13.68 & 0.01 & 13.52 & 0.01 \\
\hline
\end{tabular}}
\tablefoot{Symbols "*" and "**" pinpoint the CMD$-$ and CMD+HRD$-$old sources, respectively, as presented in Figure \ref{fig:HRD}.}
\end{center}
\label{tab:omegacam_info}
\end{table}

\renewcommand{\arraystretch}{1.2}

\begin{table*}[h!]
\caption{Cross-matched IDs and Gaia DR3 astrometric parameters for the Orion Nebula disc sample presented.}
\begin{center}
\resizebox{\textwidth}{!}{
\begin{tabular}{lccccccccc}
\hline \hline
ID & Gaia DR3 & OmegaCAM & RA & DEC & plx & pm & pmRA & pmDEC & RUWE     \\
 & & & [hh:mm:ss.ss] & [dd:mm:ss.s] & [mas] & [mas/yr] & [mas/yr] & [mas/yr] & \\ \hline
1 & 3017365914449685888 & 1042831 & 05:35:16.07 & $-$05:20:36.3  &  2.88 $\pm$ 0.16  &  4.39  &  2.14  &  3.83  &  2.69 \\
2 & 3017364887967710976 & 1021191 & 05:34:52.17 & $-$05:22:31.9 &  2.82 $\pm$ 0.73  &  4.23  &  2.67  &  3.28  &  9.97\\
3 & 3017186251673566208 & 1079659 & 05:35:57.65 & $-$05:57:18.4 &  2.60 $\pm$ 0.02  &  1.4  &  1.21  &  -0.71  &  1.1\\
4 & 3017265244711916672 & 1035737 & 05:35:08.02 & $-$05:32:44.3 &  2.54 $\pm$ 0.05  &  0.41  &  0.41  &  -0.09  &  1.08\\
5 & 3017270291298102528 & 996321 & 05:34:26.16 & $-$05:26:30.4 &  2.61 $\pm$ 0.05  &  1.96  &  1.48  &  -1.28  &  0.95\\
6 & 3017266275504061056 & 1031924 & 05:35:03.57 & $-$05:29:26.3 &  2.15 $\pm$ 0.09  &  3.63  &  3.63  &  -0.01  &  1.15\\
7 & 3209530830107709696 & 1056861 & 05:35:31.50 & $-$05:05:01.7 &  2.66 $\pm$ 0.52  &  6.26  &  3.86  &  4.93  &  21.35\\
8* & 3017367636742162176 & 1051021 & 05:35:25.23 & $-$05:15:35.8 &  2.56 $\pm$ 0.04  &  3.82  &  3.54  &  1.43  &  1.08\\
9* & 3017359184248230272 & 1053477 & 05:35:28.10 & $-$05:29:33.5 &  2.55 $\pm$ 0.05  &  1.47  &  1.45  &  -0.26  &  0.99\\
10 & 3017266962698857216 & 985345 & 05:34:13.20 & $-$05:33:53.5 &  2.66 $\pm$ 0.03  &  1.07  &  0.19  &  1.05  &  1.04\\
11 & 3017245655363965696 & 1057420 & 05:35:32.22 & -05:44:26.6  &  2.92 $\pm$ 0.66  &  9.76  &  -6.63  &  7.16  &  26.64\\
12 & 3209525263830030336 & 1014551 & 05:34:45.20 & $-$05:10:47.6 &  2.54 $\pm$ 0.04  &  2.28  &  0.96  &  -2.07  &  1.28\\
13 & 3209518598040871296 & 1019536 & 05:34:50.45 & $-$05:20:20.4 &  2.56 $\pm$ 0.04  &  2.17  &  2.11  &  0.48  &  1.09\\
14* & 3017359665283930752 & 1065824 & 05:35:41.32 & $-$05:27:50.3 &  2.83 $\pm$ 0.08  &  1.83  &  1.46  &  1.1  &  1.69\\
15 & 3017365850035788800 & 1045603 & 05:35:18.96 & $-$05:21:07.8 &  2.44 $\pm$ 0.06  &  1.02  &  0.8  &  0.64  &  0.78\\
16 & 3017359596564784128 & 1055893 & 05:35:30.48 & $-$05:28:30.6 &  2.49 $\pm$ 0.05  &  0.27  &  0.25  &  -0.1  &  1.0\\
17 & 3017360764795569792 & 1054793 & 05:35:29.34 & $-$05:25:46.3 &  2.27 $\pm$ 0.1  &  1.91  &  1.55  &  1.12  &  1.56\\
18 & 3017347089620157440 & 1061070 & 05:35:36.36 & $-$05:31:37.8 &  2.48 $\pm$ 0.06  &  2.1  &  2.07  &  0.33  &  1.03\\
19 & 3017269505319467520 & 1025573 & 05:34:56.51 & $-$05:27:51.0  &  2.49 $\pm$ 0.05  &  1.32  &  1.04  &  -0.8  &  0.93\\
20 & 3017367739821367552 & 1048863 & 05:35:22.62 & $-$05:14:11.2  &  2.49 $\pm$ 0.06  &  2.04  &  1.14  &  1.69  &  1.04\\
21* & 3017360627356915328 & 1049701 & 05:35:23.66 & $-$05:26:27.1  &  2.41 $\pm$ 0.06  &  2.75  &  2.59  &  -0.93  &  0.96\\
22 & 3017269264801309568 & 1011929 & 05:34:42.74 & $-$05:28:37.5  &  2.76 $\pm$ 0.28  &  4.38  &  3.86  &  2.08  &  1.88 \\
23 & 3017246656093604736 & 1044085 & 05:35:17.35 & $-$05:42:14.6 &  2.51 $\pm$ 0.04  &  2.15  &  1.81  &  -1.16  &  1.19\\
24* & 3017269367880519936 & 1022028 & 05:34:52.92 & $-$05:28:59.0 &  2.51 $\pm$ 0.07  &  1.09  &  -0.76  &  0.79  &  1.05\\
25 & 3017264626236630656 & 1033812 & 05:35:05.77 & $-$05:33:55.8 &  2.57 $\pm$ 0.07  &  1.3  &  1.27  &  0.29  &  0.98\\
26 & 3017360627356913280 & 1050335 & 05:35:24.46 & $-$05:26:31.5 &  2.5 $\pm$ 0.07  &  2.2  &  1.52  &  -1.6  &  0.73\\
27* & 3017367499303217280 & 1045919 & 05:35:19.29 & $-$05:16:44.7 &  2.52 $\pm$ 0.06  &  1.15  &  0.78  &  0.85  &  1.05 \\
28* & 3017360421198666624 & 1033893 & 05:35:05.93 & $-$05:27:11.1  &  2.49 $\pm$ 0.1  &  2.02  &  1.65  &  -1.16  &  0.94\\
29 & 3017360455558400896 & 1034451 & 05:35:06.60 & $-$05:26:51.0 &  2.47 $\pm$ 0.08  &  2.66  &  2.52  &  -0.85  &  0.88\\
30** & 3017245144265120384 & 1051294 & 05:35:25.50 & $-$05:45:44.8 &  2.21 $\pm$ 0.07  &  2.08  &  2.07  &  0.18  &  1.61 \\
31* & 3017269848916851840 & 1018190 & 05:34:49.07 & $-$05:26:26.6 &  2.55 $\pm$ 0.08  &  2.53  &  2.31  &  1.03  &  1.01\\
32* & 3017363582294575744 & 1034406 & 05:35:06.66 & $-$05:25:02.9 &  2.35 $\pm$ 0.16  &  1.2  &  1.03  &  0.61  &  0.89\\
33 & 3209533029130884864 & 1018691 & 05:34:49.58 & $-$05:04:59.5 &  2.51 $\pm$ 0.05  &  1.85  &  1.53  &  -1.04  &  0.99\\ 
34* & 3017249095634906496 & 1028024 & 05:34:59.05 & $-$05:44:29.8 &  2.5 $\pm$ 0.04  &  1.01  &  0.61  &  0.81  &  1.64\\
35 & 3017191646152505088 & 1040907 & 05:35:14.03 & $-$05:52:09.0 &  2.59 $\pm$ 0.02  &  1.57  &  1.44  &  0.61  &  1.08 \\
36* & 3209572096153320192 & 1051303 & 05:35:25.52 & $-$04:51:20.5 &  2.63 $\pm$ 0.05  &  1.31  &  0.42  &  -1.24  &  1.04\\
37 & 3209559486129350912 & 1056285 & 05:35:30.90 & $-$04:55:17.7 &  2.53 $\pm$ 0.02  &  2.24  &  1.9  &  -1.18  &  1.05\\
38** & 3017205317032619008 & 930757 & 05:32:49.93 & $-$06:10:45.5 &  2.63 $\pm$ 0.06  &  1.62  &  1.57  &  0.39  &  0.99 \\
39** & 3016776683592103552 & 947152 & 05:33:20.98 & $-$06:22:33.4 &  2.54 $\pm$ 0.05  &  1.28  &  1.28  &  -0.01  &  1.21\\
40 & 3209422253332724224 & 985078 & 05:34:12.88 & $-$05:28:48.0 &  2.61 $\pm$ 0.02  &  0.5  &  0.48  &  -0.12  &  1.12\\ \hline
\end{tabular}}
\tablefoot{Symbols "*" and "**" pinpoint the CMD$-$ and CMD+HRD$-$old sources, respectively, as presented in Figure \ref{fig:HRD}.} 
\end{center}
\label{tab:source_info}
\end{table*}

\FloatBarrier

\section{Best-fit models to the Balmer jump}\label{app:FRAPPE_fits}

Figure \ref{fig:all_FRAPPE_fits} shows the best fits to the X-Shooter spectra of our sample, obtained using FRAPPE (see Section. \ref{sec:FRAPPE}). For each target, we present the fit in the Balmer jump region. 


\begin{figure}[p]
    \centering
    \includegraphics[width=0.95\textwidth]{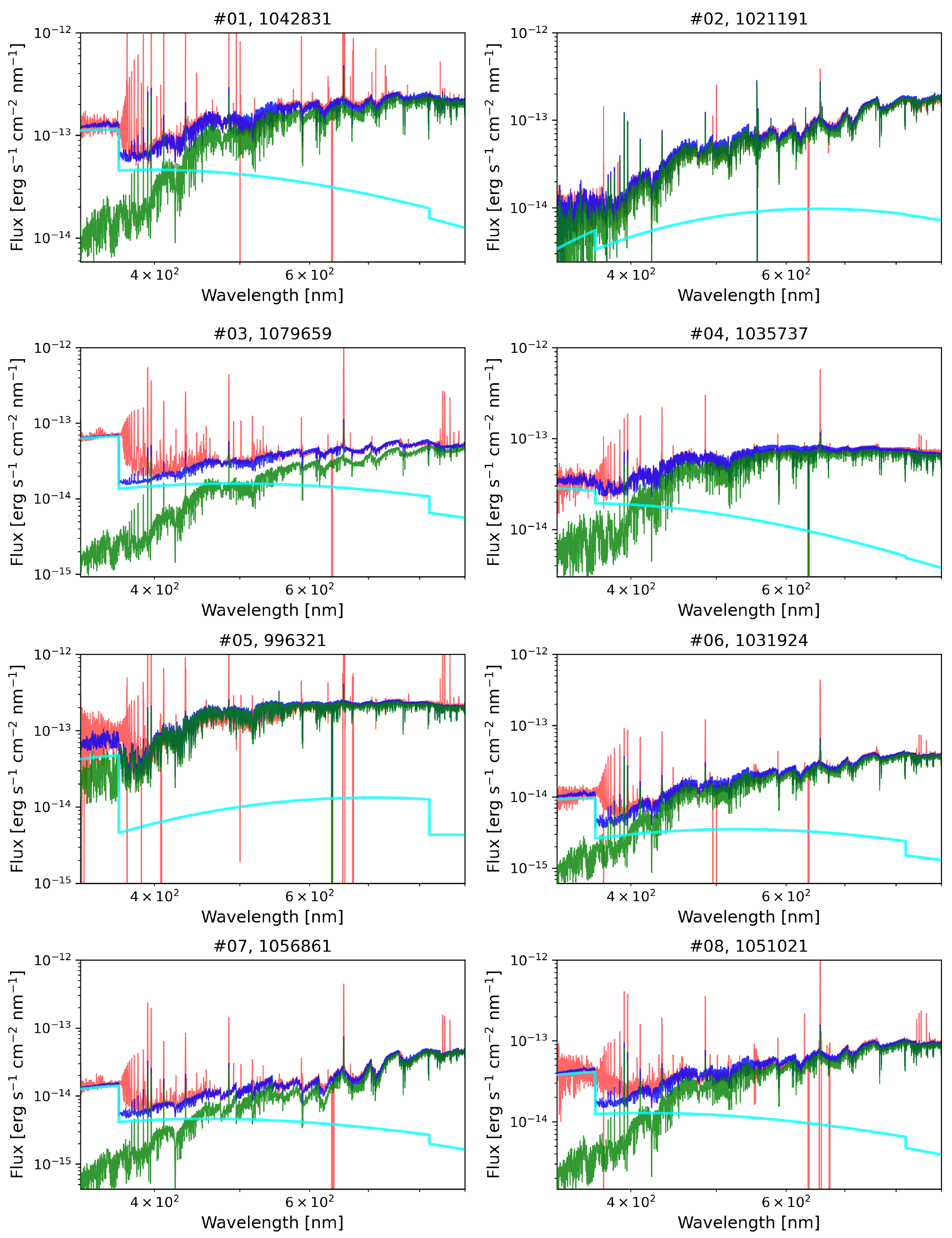}
      \caption{FRAPPE fits to the X-Shooter spectra for the Orion Nebula targets. Sources \#9, \#28, and \#32 could not be fitted with FRAPPE thus we present only their observed spectra in red. For the remaining sources, the de-reddened observed spectrum is shown in red, the best-fit Class III template in green, and the best-fit accretion slab model in cyan. The combined best-fit model (Class III $+$ slab model) is shown in dark blue. For each source both the OmegaCAM ID and the internal ID used throughout this work are indicated. All spectra were smoothed for clarity.}
      \label{fig:all_FRAPPE_fits}
\end{figure}

\begin{figure}[p]
    \ContinuedFloat
    \centering
    \includegraphics[width=0.95\linewidth]{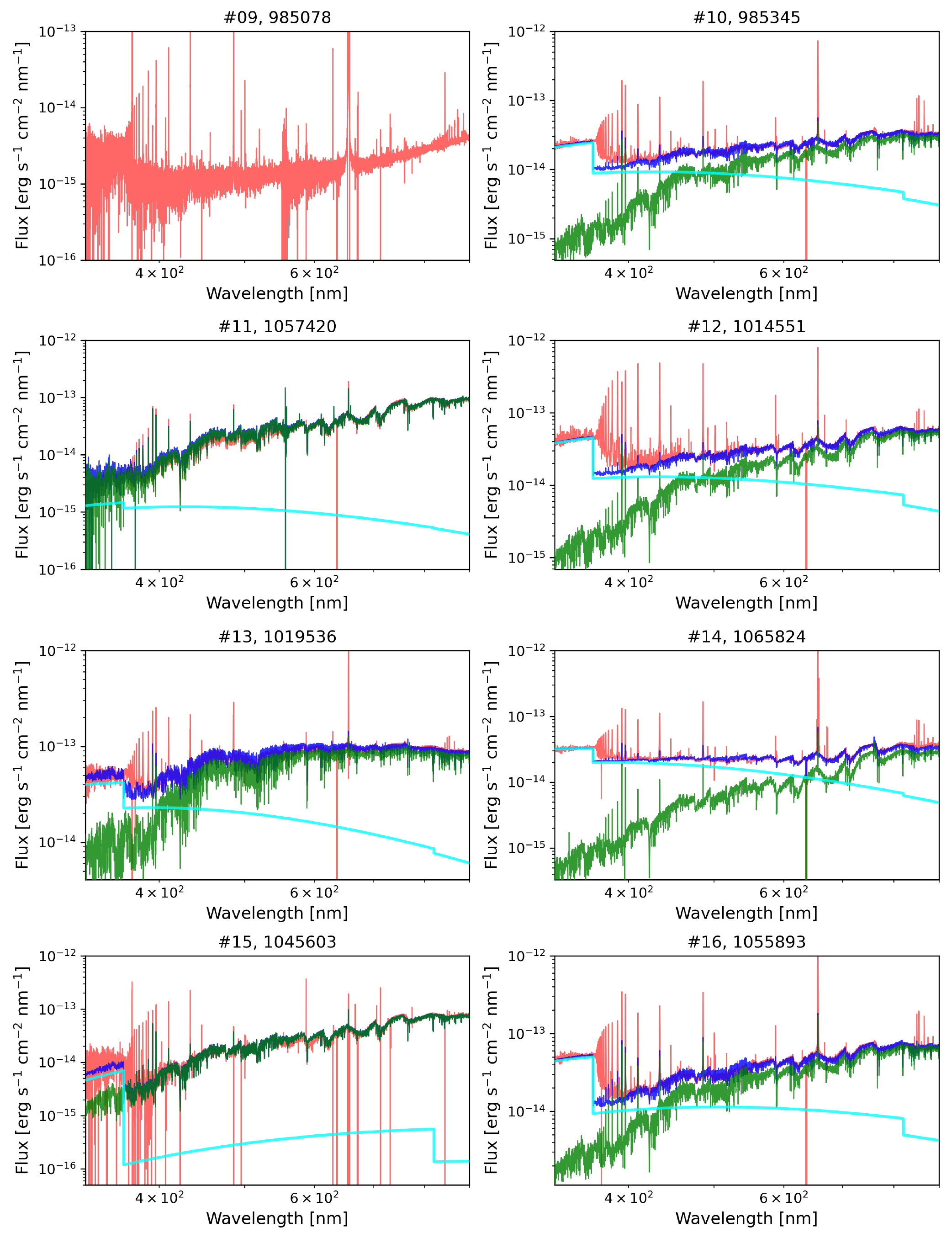}
        \caption{continued.}
        \label{fig:placeholder}
\end{figure}

\begin{figure}[p]
    \ContinuedFloat
    \centering
    \includegraphics[width=0.95\linewidth]{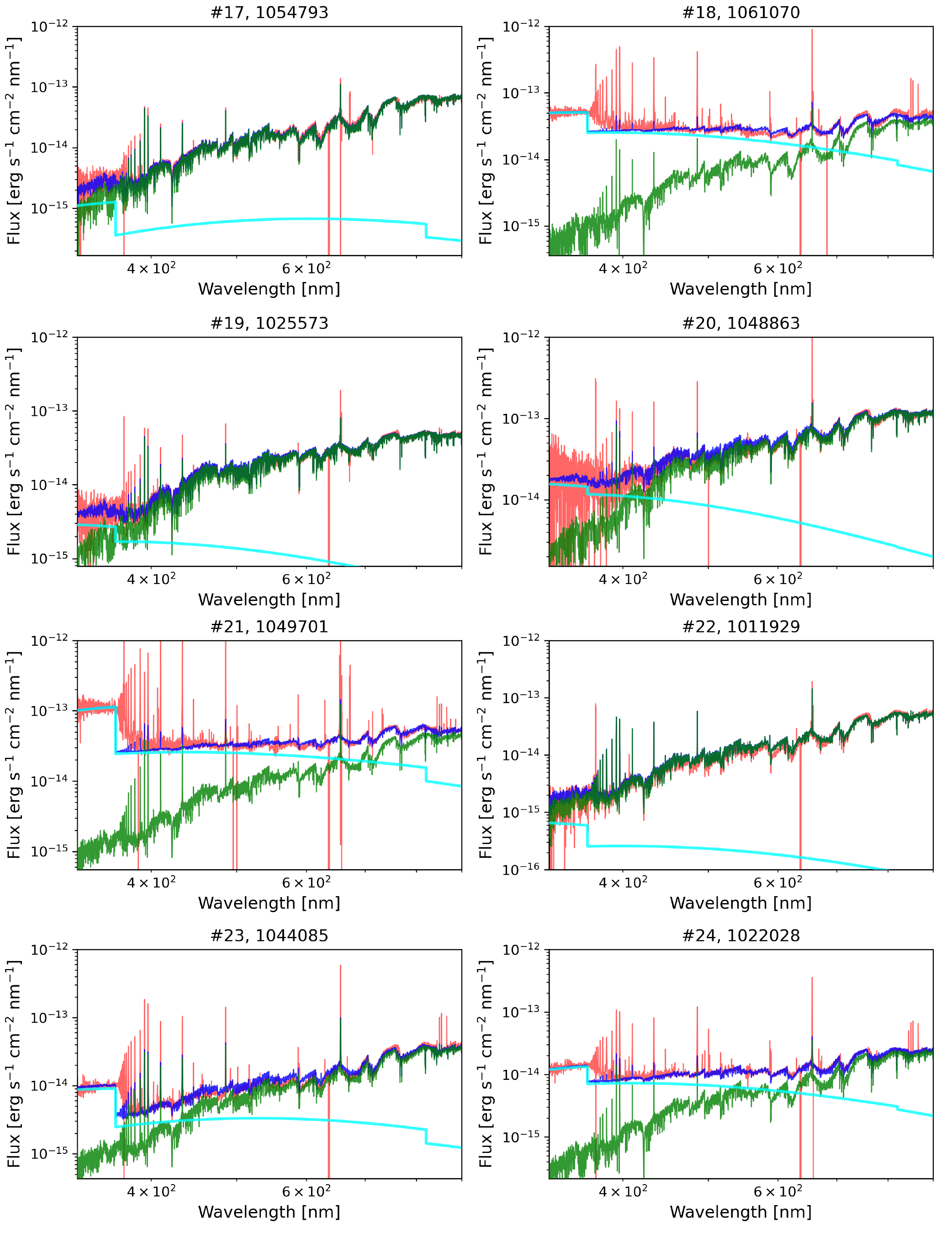}
        \caption{continued.}
        \label{fig:placeholder}
\end{figure}

\begin{figure}[p]
    \ContinuedFloat
    \centering
    \includegraphics[width=0.95\linewidth]{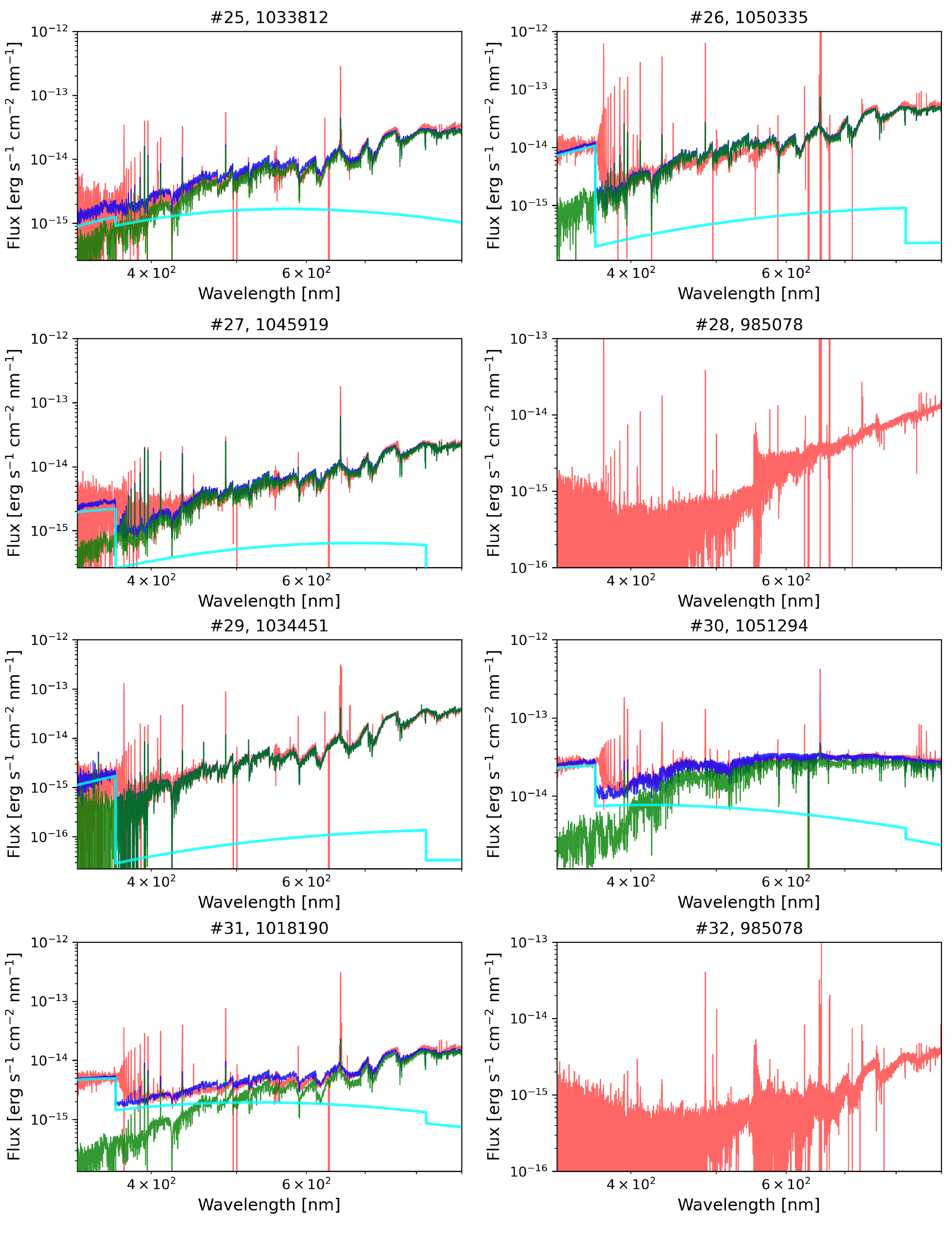}
        \caption{continued.}
        \label{fig:placeholder}
\end{figure}

\begin{figure}[p]
    \ContinuedFloat
    \centering
    \includegraphics[width=0.95\linewidth]{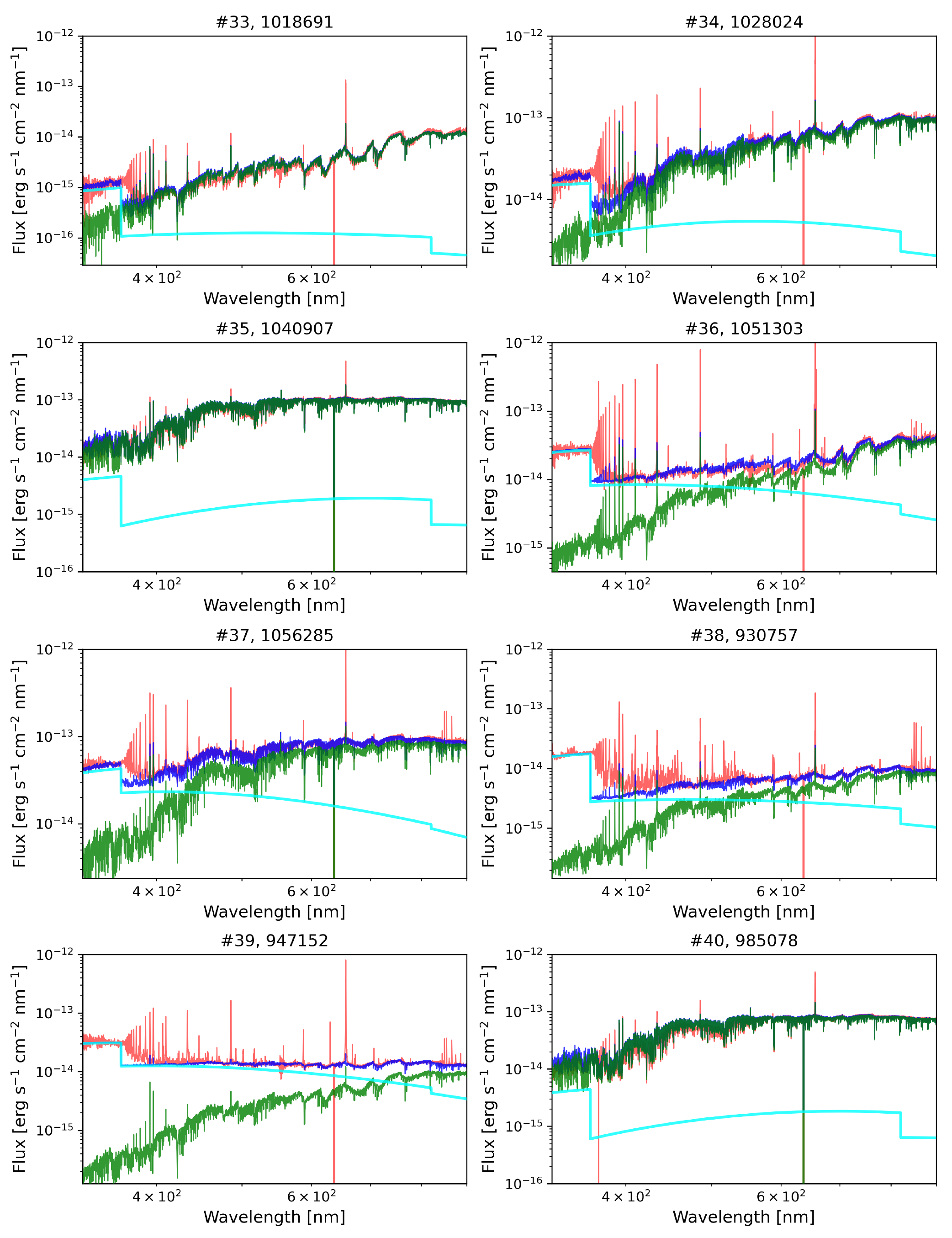}
        \caption{continued.}
        \label{fig:placeholder}
\end{figure}

\FloatBarrier

\section{Lithium in entire X-Shooter sample}

\begin{figure*}[h]
\centering
    \includegraphics[width=0.7\textwidth]{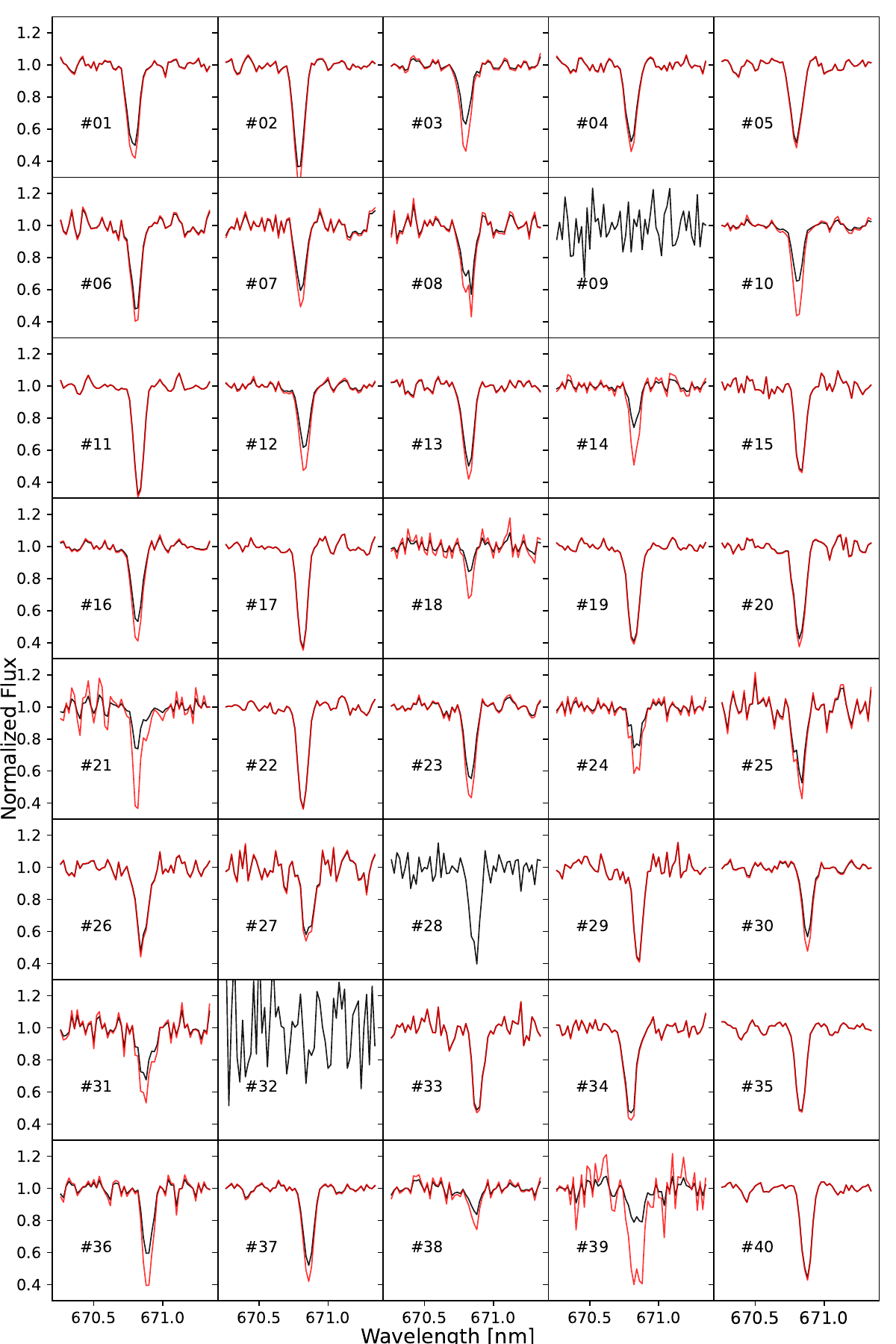}
\caption{Observed and veiling-corrected Li I $\lambda$670.8 nm absorption in our sample (in black and red, respectively). The objects without an associated veiling-corrected line are those for which FRAPPE could not be implemented. Each spectrum is labeled with the corresponding object ID, shown at the bottom of each panel. All spectra are normalized to the local continuum near the line.}
\label{fig:lithium_sample}
\end{figure*}

In Figure \ref{fig:lithium_sample}, we show the Li$\lambda$670.8 $\mu$m feature of the entire X-Shooter sample. 

\end{appendix}

\end{document}